\documentclass[preprint]{JHEP} 

\usepackage{epsfig}                     

\newbox\mybox
\newcommand\fverb{\setbox\mybox=\hbox\bgroup\verb}
\newcommand\fverbdo{\egroup\medskip\noindent\fbox{\unhbox\mybox}\ }
\newcommand\fverbit{\egroup\item[\fbox{\unhbox\mybox}]}

\font\beeg=cmr17 scaled 1600            
\newcommand\init[1]{\setbox\mybox=\hbox{{\beeg #1}~}%
                   \noindent\global\hangindent=\wd\mybox\global\hangafter-2%
                   \sc\smash{\llap {\lower 13.2pt \box\mybox}}}

\newcommand{\be}{\begin{equation}}
\newcommand{\ee}{\end{equation}}
\newcommand{\lsim}   {\mathrel{\mathop{\kern 0pt \rlap
  {\raise.2ex\hbox{$<$}}}
  \lower.9ex\hbox{\kern-.190em $\sim$}}}
\newcommand{\gsim}   {\mathrel{\mathop{\kern 0pt \rlap
  {\raise.2ex\hbox{$>$}}}
  \lower.9ex\hbox{\kern-.190em $\sim$}}}

\def\be{\begin{equation}}
\def\ee{\end{equation}}
\def\ba{\begin{eqnarray}}
\def\ea{\end{eqnarray}}

\def\e{{\rm e}}
\def\ap{\approx}

\def\Eb{E_{\rm b}}
\def\Ee{\langle E_{\bar\nu_{\rm e}}\rangle}
\def\Eh{\langle E_{\bar\nu_{\rm h}}\rangle}
\def\dm{\Delta m^2}

\def\L{{\mathcal L}}
\def\theta{\vartheta}
\def\epsilon{\varepsilon}

\title{Large lepton mixing and supernova 1987A}

\author{M. Kachelrie{\ss}\\
        {\it\small TH Division, CERN, Geneva} }

      \author{R. Tom\`as and J.W.F. Valle\\
        {\it \small Instituto de Fisica Corpuscular -- C.S.I.C.
          -- Universitat de Val\`encia  \\
          Ed. de Institutos de Paterna -- Apartado de Correos 22085 -
          46071 Val\`encia, Spain}}

\preprint{CERN-TH 2000-363\\ IFIC/00-77}

\abstract{
  
  We reconsider the impact of $\bar\nu_e \leftrightarrow
  \bar\nu_{\mu,\tau}$ neutrino oscillations on the observed
  $\bar\nu_e$ signal of supernova SN 1987A. Performing a
  maximum-likelihood analysis using as fit parameters the released
  binding energy $\Eb$ and the average neutrino energy $\Ee$, we find
  as previous analyses that $\bar\nu_e \leftrightarrow
  \bar\nu_{\mu,\tau}$ oscillations with large mixing angles have lower
  best-fit values for $\Ee$ than small-mixing angle (SMA)
  oscillations. Moreover, the inferred value of $\Ee$ is already in
  the SMA case lower than those found in simulations.  This apparent
  conflict has been interpreted as evidence against the large mixing
  oscillation solutions to the solar neutrino problem.  In order to
  quantify the degree to which the experimental data favour the SMA
  over the large mixing solutions we use their likelihood ratios as
  well as a Kolmogorov-Smirnov test. We find within the range of SN
  parameters predicted by simulations regions in which the LMA-MSW
  solution is either only marginally disfavoured or favoured compared
  to the SMA-MSW solution. We conclude therefore that the LMA-MSW
  solution is not in conflict with the current understanding of SN
  physics. In contrast, the vacuum oscillation and the LOW solutions
  to the solar neutrino problem can be excluded at the $4\sigma$ level
  for most of the SN parameter ranges found in simulations.  Only a
  marginal region with low values of $\Ee$, $\langle
  E_{\bar\nu_{\mu,\tau}} \rangle$ and $\Eb$ is left over, in which
  these oscillation solutions can be reconciled with the neutrino
  signal of SN~1987A.}


\begin{document} 

\section{Introduction}
The impact of neutrino properties beyond the standard model on the
observed neutrino signal~\cite{obs} of supernova (SN) SN~1987A has
been discussed in numerous works during the last 13 years. Most of the
early works discussed either mass limits for $\nu_e$ or the impact of
matter-induced oscillations on the prompt $\nu_e$ burst~\cite{alt}. In
the latter case the MSW effect in the SN envelope would have rendered
the prompt $\nu_e$ burst unobservable for a large area of mixing
parameters assuming a normal mass hierarchy. On the other hand, the
main neutrino signal of SN~1987A which is the $\bar\nu_e$ pulse
detected by the reaction $\bar\nu_e+p \to n+e^+$ is then influenced
only for large mixing angles (LMA) by oscillations.  It was
argued~\cite{ssb,jnr} that LMA neutrino oscillations would increase
the discrepancy between the observed $\bar\nu_e$ spectrum and the one
predicted by SN simulations~\cite{Bu88,SN-sim,ja93,Bu00}.  Moreover,
the theoretical prejudice was until recently that lepton mixing is,
similar to quark mixing, small.  Therefore, the small mixing angle
SMA-MSW solution\footnote{This is, for the purposes of the present
  SN analysis, equivalent to the no-oscillation hypothesis.} to the
solar neutrino problem seemed to be favoured both by the neutrino
signal of SN~1987A and by theoretical arguments.

Meanwhile, the LMA-MSW solution to the solar neutrino problem has
become the best oscillation solution to the solar neutrino
problem~\cite{global00}, while Superkamiokande atmospheric neutrino
data~\cite{SK} strongly indicate the need for $\nu_\mu \to \nu_{\tau}$
oscillations with maximal or nearly maximal mixing~\cite{global00}.
Such approximate bi-maximal neutrino mixing patterns~\cite{bimax98}
can actually arise either in theoretical models based on unification
ideas~\cite{bimaxmodtb00} or in ``bottom-up'' models, such as that of
Ref.~\cite{Hirsch:2000ef}.
It is therefore interesting to reconsider how serious the discrepancy
between the observed and the predicted $\bar\nu_e$ spectra of SN~1987A
should be taken.

The main goal of this work is not to obtain those values of the
neutrino oscillation parameters, $\tan^2\theta$ and $\dm$, that fit
best the neutrino signal from SN~1987A.  Instead we use the best-fit
points found from solar neutrino data for three [SMA-MSW, LMA-MSW, and
vacuum oscillations (VO)] of the four allowed solutions as input and
try then to assess to what extend the LMA solutions are (dis-)
favoured compared to SMA assuming certain astrophysical parameters.
The case of the LOW solution which extends as only one also into the
so-called dark side ($\theta>\pi/4$) of the solar neutrino
problem~\cite{dark} will be discussed separately. There, we are mainly
interested in the question if the dark side is compatible with the
neutrino signal from SN~1987A~\cite{ssb,inv}.  In all cases, we use as
measure for the degree to which the experimental data favour one of
the allowed solutions over the others their likelihood ratios as well
as a Kolmogorov-Smirnov test.

\begin{table}
\begin{center}
\begin{tabular}{|c|c|c|c|} 
\hline 
                    & SMA & LMA & VO  \\ \hline
$\Delta m^2$/eV$^2$ & $5.0\times 10^{-6} $  & $3.2\times 10^{-5} $  
            & $1.0\times 10^{-10} $   \\ 
$\tan^2\theta$ & 0.00058 & 0.33 & 0.52 \\ \hline             
\end{tabular}
\end{center}
\caption{\label{SNP}
Best-fit values for the mass difference squared $\dm$ and the mixing
angle for different oscillation solutions to the solar neutrino
problem, from Ref.~\cite{global00}.}
\end{table}

Our main result is that there is a region in the space of
astrophysical parameters (average neutrino energies $\langle
E_i\rangle$ and released binding energy $\Eb$) predicted by SN
simulations in which the LMA-MSW solution is either only marginally
disfavoured or favoured compared to the SMA-MSW solution. For the range
of parameters compiled in Ref.~\cite{ja93},
\ba   \label{box1}
  &&  14~{\rm MeV}  \leq \Ee  \leq 17~{\rm MeV}    \\
  &&  24~{\rm MeV}  \leq \langle E_{\bar\nu_{\mu,\tau}} \rangle 
                      \leq 27~{\rm MeV}    \\
       \label{box3}
  &&  1.5\times 10^{53}~{\rm erg}\leq \Eb\leq 4.5\times 10^{53}~{\rm erg}, 
\ea
the probability that the LMA-MSW hypothesis is compatible with the
detected neutrino signal is extremely varying.
We find that for large values of $\Ee$, $\langle
E_{\bar\nu_{\mu,\tau}} \rangle$ and $\Eb$ all three large mixing
solutions solutions are practically excluded when compared to the
SMA-MSW solution.
However the probability that the LMA-MSW hypothesis is compatible with
the neutrino signal rises above 10\% for values of $\Ee$, $\langle
E_{\bar\nu_{\mu,\tau}} \rangle$ and $\Eb$ near the lower end of
(\ref{box1}-\ref{box3}).  
%
%
Moreover, it was recently argued that the average neutrino energies 
are smaller than earlier believed. For example, nucleon-nucleon 
bremsstrahlung that has not been included into the earlier supernova 
codes tends to soften the $\bar\nu_{\mu,\tau}$ spectrum~\cite{nn}, 
a trend which has been 
confirmed by a recent simulation~\cite{Bu00}. This simulation found also 
for $\Ee$ a rather low value, $\Ee \ap 12$~MeV. Since this simulation
has not aimed at a self-consistent treatment of the neutrino spectra,
the values found therein should be taken only as an indication. 
If $\Ee$ is indeed as low as found in Ref.~\cite{Bu00}, 
the LMA-MSW solution can be {\em favoured\/} compared to SMA-MSW.

In contrast, the VO and the LOW solutions to the solar neutrino
problem can be excluded at the $4\sigma$ level for most of the range
of SN parameters found in simulations, and low neutrino energies,
$\Ee\lsim 12$~MeV and $\langle E_{\bar\nu_{\mu,\tau}} \rangle \lsim
17$~MeV, combined with low $\Eb$ are required to reconcile these
oscillation solutions with the neutrino signal of SN~1987A.

\section{Neutrino fluences and oscillations}
Massive stars with mass $M \gsim 8~M_\odot$ end their
lives in spectacular type-II supernova outbursts, releasing almost all
their gravitational binding energy via neutrino
emission~\cite{reviews}. Numerical simulations~\cite{Bu88,SN-sim,ja93}
as well as analytic considerations~\cite{ja95} show that this energy
is approximately equipartioned between the three neutrino flavours.
The instantaneous neutrino spectra found in simulations are generally
pinched, i.e. their low- and high-energy parts are suppressed relative
to a Maxwell-Boltzmann distribution.  The pinching of the
instantaneous neutrino spectra is however compensated by the
superposition of different spectra with decreasing temperatures.
Moreover, we have found that the likelihood function depends only
weakly on how strongly the spectra are pinched. We present therefore
only results using Maxwell-Boltzmann distributions for the
time-averaged spectra of the neutrinos. Finally, we assume that the SN
emits the same amount of energy in all neutrino flavours.

We perform our analysis in the framework of two-neutrino oscillations
$\bar\nu_e \leftrightarrow \bar\nu_h
=\cos\phi\:\bar\nu_\mu+\sin\phi\:\bar\nu_\tau$, which is motivated
both by detailed fits of the atmospheric neutrino
anomaly~\cite{global00}, but also by the results of the Chooz
experiment~\cite{Bemporad:1999}.  Since the energy spectra of
$\bar\nu_\mu$ and $\bar\nu_\tau$ are identical (up to ${\cal
  O}(E^4G_F^2)$ effects), our results do not depend on $\phi$.
The probability of a $\bar\nu_e$ to arrive at the surface of the Earth
can be written as an incoherent sum of probabilities,
\be
\label{Pee}
  P_{\bar e \bar e} = 
  P^S_{\bar e1} P^E_{1\bar e} + P^S_{\bar e2} P^E_{2\bar e} =  
  \left(1-P_c\right)\cos^2\theta + P_c\sin^2\theta \,,
\ee
where $P^S_{\bar ei}$ denotes the probability that a $\bar\nu_e$ leaves the
star as mass eigenstate $\bar\nu_i$ and $P^E_{i\bar e}$ the probability
that $\bar\nu_i$ is detected as $\bar\nu_e$. 
The interference terms can be safely neglected for all $\Delta m^2$ 
of interest because of the spread of the neutrino wave packets on 
their way to the Earth.
Deviations from an adiabatic
evolution of the neutrino states in the SN envelope are characterised
by the crossing probability $P_c$. We use as approximation for $P_c$
the expression valid for an exponential density profile,
\be  \label{exp}
P_c = \frac{\e^{-\gamma\cos^2\theta}-\e^{-\gamma}}{1-\e^{-\gamma}} \,,
\ee
where the adiabaticity parameter
\be 
 \gamma = \frac{4\pi^2 l_\rho}{l_{\rm osc}} \ap
          10^9 \; \left( \frac{\dm}{{\rm eV}^2} \right) 
                  \left( \frac{\rm MeV}{E} \right)
\ee
is the ratio of density scale height $l_\rho$ and neutrino oscillation
length $l_{\rm osc}$ in vacuum. The expression~(\ref{exp}) has in
contrast to the one for a linear density profile the correct
non-adiabatic limit, $P_c\to\sin^2\theta$, and describes also for
$\gamma>0$ quite accurately the behaviour of $P_c$ for a profile
typical for a SN envelope, $N_e\propto r^{-3}$~\cite{Ku89}.
Furthermore, for the case of nearly maximal mixing in which we are
especially interested $P_{\bar e\bar e}=1/2+{\cal O}(\theta^2)$.

The neutrinos from SN~1987A had to cross the mantle of the Earth
before they reached the Kam and IMB detectors. Possible matter effects
in the Earth can be approximated by a box potential, i.e. by replacing
$P^E_{i\bar e}=|\langle\bar\nu_i|\bar\nu_e\rangle|^2$ with~\cite{ssb}
\ba 
 \cos^2 \theta &\to& \cos^2 \theta + 
     \sin 2\theta^\prime \sin (2\theta-2\theta^\prime) 
     \sin^2 (\pi d/l_{\rm osc}^\prime) \\
 \sin^2 \theta &\to& \sin^2 \theta - 
     \sin 2\theta^\prime \sin(2\theta-2\theta^\prime) 
     \sin^2 (\pi d/l_{\rm osc}^\prime) \,,
\label{matter}
\ea
where primed quantities are evaluated in matter.
The distance $d$ traveled through the mantle by the neutrinos and the
average density $\rho$ are different for the Kamiokande and IMB detectors.
For Kamiokande we use $d=3900$~km and $\rho=3.4$~g/cm$^3$ and for
IMB $d=8400$~km and $\rho=4.6$~g/cm$^3$~\cite{ssb}. 

Let us now briefly consider the two extreme limits of the general
expressions Eqs.~(\ref{Pee}-\ref{matter}). For small enough $\dm$,
neutrino states evolve in the SN envelope strongly non-adiabatically,
$\gamma\to 0$ and $P_c\to\sin^2\theta$. Moreover the matter effects in
the Earth can be neglected and thus the survival probability becomes
identical to the vacuum oscillation probability, $P_{\bar e\bar
  e}\to\cos^4\theta + \sin^4\theta$.  In the opposite limit, the
evolution of the neutrino states in the SN envelope is completely
adiabatic, $\gamma\to\infty$ and $P_c\to 0$.  Thus
\be 
 P_{\bar e\bar e} = P^S_{\bar e1} P^E_{1\bar e} =  \cos^2\theta +
     \sin 2\theta^\prime \sin (2\theta-2\theta^\prime) 
     \sin^2 (\pi d/l_{\rm osc}^\prime) \,.
\ee
Hence, Eqs.~(\ref{Pee}-\ref{exp}) describe correctly the survival
probability of $\bar\nu_e$ neutrinos in the region of parameters 
favoured by the LMA-MSW and the VO solution to the solar neutrino
problem and interpolate smoothly for intermediate values of
$\dm$ and $\theta$. 
The $\bar{\nu}_e$ fluence arriving at the detectors is then 
\be
F_{\bar{\nu}_e} = P_{\bar{e}\bar{e}}F_{\bar{\nu}_e}^0 +
(1-P_{\bar{e}\bar{e}})F_{\bar{\nu}_h}^0
\ee
where $F_{\bar{\nu}}^0$ stands for the time-integrated flux
of neutrinos emitted by the SN.

\section{Likelihood analysis}
The maximum-likelihood method is a particularly well-suited tool for
problems like the one at hand, where we want to extract the maximal
possible information from only 19 neutrino events. We test with the
likelihood function 
\be
 {\cal L}(\alpha) \propto \exp\left(-\int n(E,\alpha) \; dE \right) 
               \prod_{i=1}^{N_{\rm obs}} n(E_i,\alpha)
\ee
the hypothesis that a prescribed neutrino fluence
$F_{\bar\nu_e}(E_{\bar\nu_e},\alpha)$ leads to the observed
experimental data $E_i$ with probability distribution $n(E,\alpha)$.
The maximisation of ${\cal L}(\alpha)$ gives an estimate of the values
$\alpha_\ast=\{\tan^2\theta,\dm,\Eb,\Ee,\Eh,\ldots\}$ which best
represent the data set $E_i$.  The confidence region around
$\alpha_\ast$ which contains the true value of $\alpha$ with a
specified probability $\beta$ for $k$ fit parameters is given by
\be
\label{likelihood.test}
 \ln{\cal L}(\alpha_\ast)-\ln{\cal L}(\alpha) \leq
 \frac{1}{2}\chi^2_{\beta;k} \,, 
\ee
if ${\cal L}(\alpha)$ is assumed to be Gaussian near its extrema. 

Apart from parameter estimation given a specific hypothesis, the
likelihood analysis can also be used to decide which hypothesis fits
better the experimental data.  In this case, the ratio of the
likelihood functions for the different hypotheses is a useful
estimator. We will present the best-fit values for a given oscillation
hypothesis in Sec.~\ref{para}, while we test which hypothesis fits
better the data given a certain range of parameters in
Sec.~\ref{hypo}.

The connection between the observed data set $E_i$ and the emitted
neutrino fluence $F_{\bar\nu_e}(E_{\bar\nu_e},\alpha)$ has been
described already in detail in the literature. We follow here closely
Ref.~\cite{jnr}, but use for the detector efficiencies the fit
functions given by Burrows~\cite{Bu88}.

\subsection{Best-fit values and confidence regions}
\label{para}
In this subsection we review the best-fit values and confidence
regions obtained from our likelihood analysis for the different
oscillation hypotheses.  Since the likelihood function we use is
identical to the one of Ref.~\cite{jnr}, our results differ only
slightly owing to small shifts of the best-fit values of the solar
neutrino data indicated by the most recent analysis~\cite{global00}.
Therefore we present only briefly the main points of this analysis.

In the following, the region of SN parameters (\ref{box1}-\ref{box3})
given in Ref.~\cite{ja93} corresponds always to the cross hatched
region in the figures, while the lower values of $\Ee$ found in
Ref.~\cite{Bu00} are shown as hatched region.  The best-fit values
obtained for the three different oscillation hypotheses are
summarized in Table~\ref{fit}.

The presence of $\bar\nu_e \leftrightarrow \bar\nu_{\mu,\tau}$
neutrino oscillations can affect the average energy of $\bar\nu_e$.
We have parameterised the $\bar\nu_h$ energy as $\Eh=\tau\Ee$.
In Fig.~\ref{contour-LMA}, we show the 95.4\% C.L. likelihood contour
for the LMA-MSW oscillation hypothesis in the $\Eb$-$\Ee$ plane. The
neutrino oscillation parameters are given in Table~\ref{SNP} and the
average energy of $\bar\nu_h$ is specified by the parameter $\tau$.
The case $\tau=1$ corresponds to the Standard Model or SMA-MSW
oscillations, because in both cases oscillations can be neglected
altogether.  The best-fit point is shifted to smaller values of $\Ee$
for increasing values of $\tau$, and the 95.4\% C.L. likelihood
contour includes a large portion of the hatched region for $\tau=1.4$
and 1, but only touches it for $\tau=1.7$.  Parts of the cross-hatched
region (\ref{box1}-\ref{box3}) are however included by none of the
oscillation hypotheses, and only in the case of the SMA-MSW solution
the 95.4\% C.L. likelihood contour touches nearly the boundary of
(\ref{box1}-\ref{box3}).

\FIGURE[h]{\epsfig{file=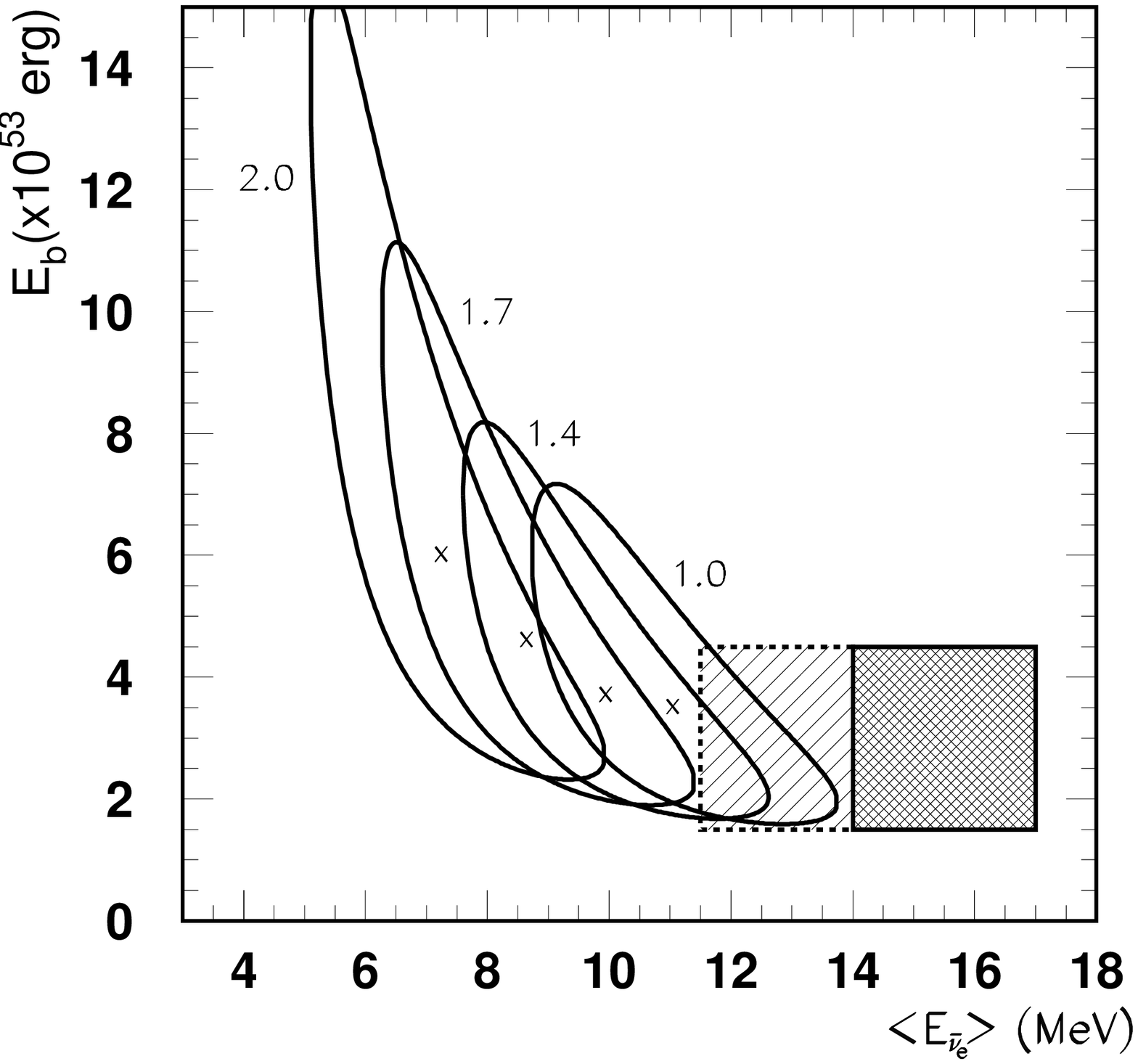,height=8.5cm,width=12cm}
           \caption{\label{contour-LMA}
             Contours of 95.4\% C.L. likelihood for the LMA-MSW
             oscillation hypothesis as function of $\Eb$ and $\Ee$ for
             $\tau=1, 1.4, 1.7$ and 2. The best-fit points are marked
             by crosses.}}
         
         In Fig.~\ref{contour-VO}, we show the same plot for the VO
         hypothesis.  The oscillation parameters are again given in
         Table~\ref{SNP}.  The shift of the best-fit point towards
         smaller values of $\Ee$ is now more pronounced and already
         for $\tau=1.4$ the 95.4\% C.L. likelihood contour does not
         include $\Ee\ap 12$~MeV. If the VO solution is really
         realized in Nature, then the SN~1987A data would indeed
         strongly favour smaller neutrino temperatures as it is
         usually assumed.

\FIGURE[h]{\epsfig{file=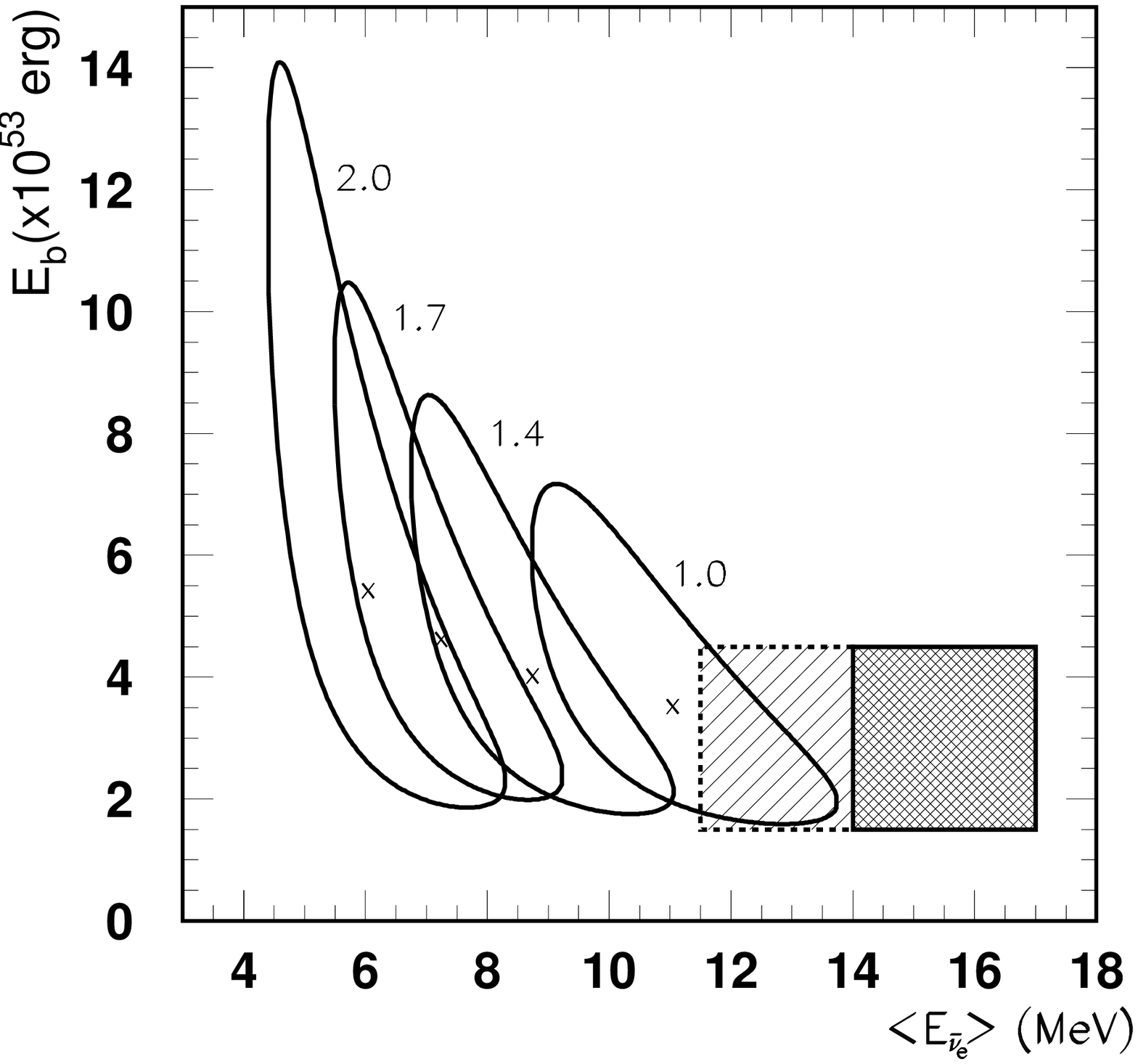,height=8.5cm,width=12cm}
           \caption{\label{contour-VO}
             Contours of 95.4\% C.L. likelihood for the VO oscillation
             hypothesis as function of $\Eb$ and $\Ee$ for
             $\tau=1,~1.4,~1.7$ and 2. The best-fit points are marked
             by crosses.}}

\begin{table}
\begin{center}
\begin{tabular}{|l|c|c|c|c|c|c|c|} 
\hline 
   &  \multicolumn{2}{c|}{Kam} &
   \multicolumn{2}{c|}{IMB} 
 & \multicolumn{3}{c|}{joint}\\ \hline
   & $\langle E_{\bar{\nu}_e}\rangle$ & $\Eb$ & 
    $\langle E_{\bar{\nu}_e}\rangle$ & $\Eb$ & 
    $\langle E_{\bar{\nu}_e}\rangle$ & $\Eb$ & $\Delta \chi ^2$ \\ \hline
 SMA             & 7.8 & 5.3 & 11.3 & 4.8 & 10.9 & 3.4 & 1.9  \\ \hline
 LMA, $\tau=1.4$ & 6.6 & 6.6 & 9.9 & 5.8 & 9.8 & 3.6 &   2.0  \\ \hline
 LMA, $\tau=1.7$ & 5.6 & 8.8 & 7.2 & 16.2 & 8.5 & 4.5 &  1.6   \\ \hline
 LMA, $\tau=2$   & 4.8 & 10.8 & 5.9 & 23.6 & 7.1 & 5.9 & 0.8  \\ \hline
 VO,~ $\tau=1.4$ & 6.0 & 6.6 & 8.4 & 7.3 & 8.6 & 3.9 &  1.2   \\ \hline
 VO,~  $\tau=1.7$ & 5.1 & 7.3 & 6.9 & 8.3 & 7.1 & 4.5 & 0.4  \\ \hline
 VO,~  $\tau=2$   & 4.2 & 8.7 & 5.7 & 9.9 & 5.9 & 5.3 & 0  \\ \hline
\end{tabular}
\end{center}
\caption{\label{fit} Best fit values for $\Ee$/MeV and 
   $\Eb/10^{53}$~erg; both for a separate and a joint analysis of the
   two experiments. The difference $\Delta\chi^2$ of the local minima
   to the global minimum (VO solution with $\tau=2$) is also given.}
\end{table}

The compatibility of the two data sets of Kamiokande and IMB has been
already discussed extensively in the literature.  Recently, it has
been speculated that the LMA-MSW solution can improve the agreement
between the two experiments~\cite{Lu00}. We have therefore analysed
the two data sets for the SMA-MSW and the LMA-MSW case also
separately. The separate contours of constant likelihood of the Kam
and IMB data sets are shown in Fig.~\ref{comparison} for SMA and
LMA-MSW oscillations with $\tau=1.4$ and $\tau=1.7$, respectively.  As
an intuitive measure for the agreement of the two experiments we give
in Table~\ref{abs} the absolute size of the intersection of fixed
likelihood regions. If one allows for arbitrary values of $\Eb$, the
overlapping area of both the 90\% and the 95\% C.L. regions increases
considerably for increasing $\tau$, as can be seen from Table
\ref{abs}.  Since however the likelihood contours of the individual
experiments extend up to unrealistic large values of $\Eb$, we present
results not only for the case that $\Eb$ can float arbitrarily but
also for the case that $\Eb$ is restricted to lie in the band
$1.5\times 10^{53}~{\rm erg}\leq\Eb\leq 4.5\times 10^{53}$~erg.  Then,
the absolute size of the intersection of the two experiments
diminishes slightly for increasing $\tau$. The same is true if one
uses the relative size defined as fraction of the intersection with
the union of the areas. We conclude therefore that the LMA-MSW
solution with the current best-fit values for $\dm$ and $\theta$ does
not improve significantly the compatibility of the two data sets. The
compatibility might improve however for a judicious tuning of $\dm$
and $\theta$~\cite{Lu00}.
Finally, we want to comment on the very large values of $\Eb$ found to
be compatible with the IMB data set (cf. Fig.~\ref{comparison}).  Both
the energy dependence of the cross-section $\sigma(\bar\nu_ep\to
ne^+)\propto E_\nu^2$ and of the detector efficiency made the IMB
detector blind for $\bar\nu_e$ with $E_\nu\lsim 20$~MeV and then
increasingly sensitive to $\bar\nu_e$ with higher energies. Therefore,
a decrease in $\Ee$ can be compensated by an increase in $\Eb$ such
that enough neutrinos from the Boltzmann suppressed high-energy tail
of the distribution are still detected, while the main part of the
emitted neutrino signal is not seen. Thus $\Eb$ and $\Ee$ are,
especially in the IMB data set, strongly correlated and the shape of
the likelihood contours is very distorted.
\begin{table}
\begin{center}
\begin{tabular}{|l|c|c|c|c|} 
\hline 
  &  \multicolumn{2}{c|}{$\Eb$ restricted} &
  \multicolumn{2}{c|}{$\Eb$ free} \\ \hline
  & 90\% C.L. & 95.4\% C.L.   & 90\% C.L. & 95.4\% C.L.  \\ \hline
SMA             &    6.5 & 15 &   8.5 & 22 \\ \hline
LMA, $\tau=1.4$ &    5.7 & 14 &   8.8 & 24 \\ \hline
LMA, $\tau=1.7$ &    5.1 & 13 &   12  & 31 \\ \hline
LMA, $\tau=2$   &    4.4 & 12 &   19  & 45 \\ \hline
\end{tabular}
\end{center}
\caption{\label{abs}
         Absolute size (in arbitrary units) of the overlapping part of
         the 90\% and 95.4\% confidence regions of the two data sets.}
\end{table}

\FIGURE[h]{
\vspace* {-1.3cm}
           \epsfig{file=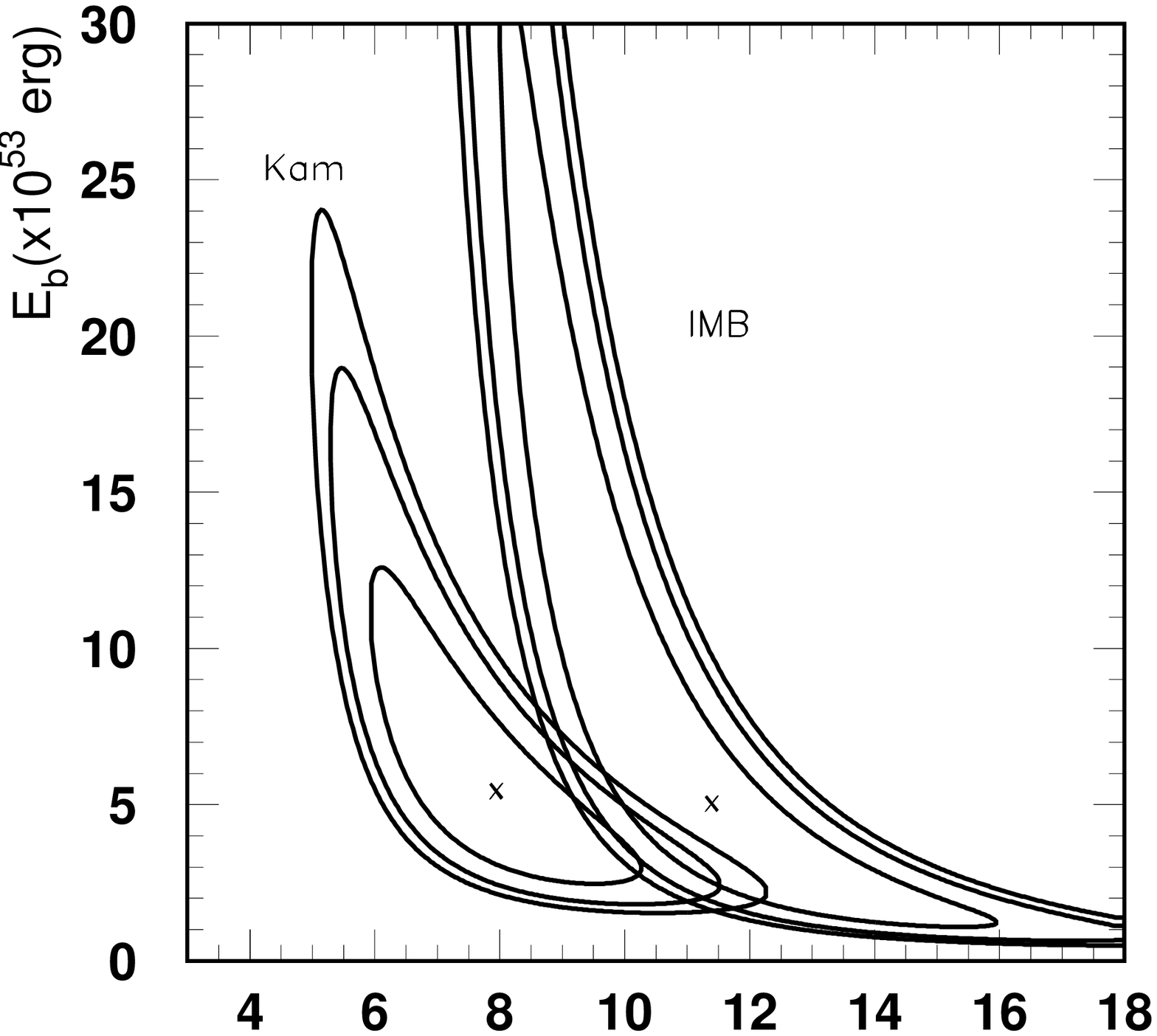,height=8cm,width=12cm,angle=0}  
\vspace* {-1.3cm}
           \epsfig{file=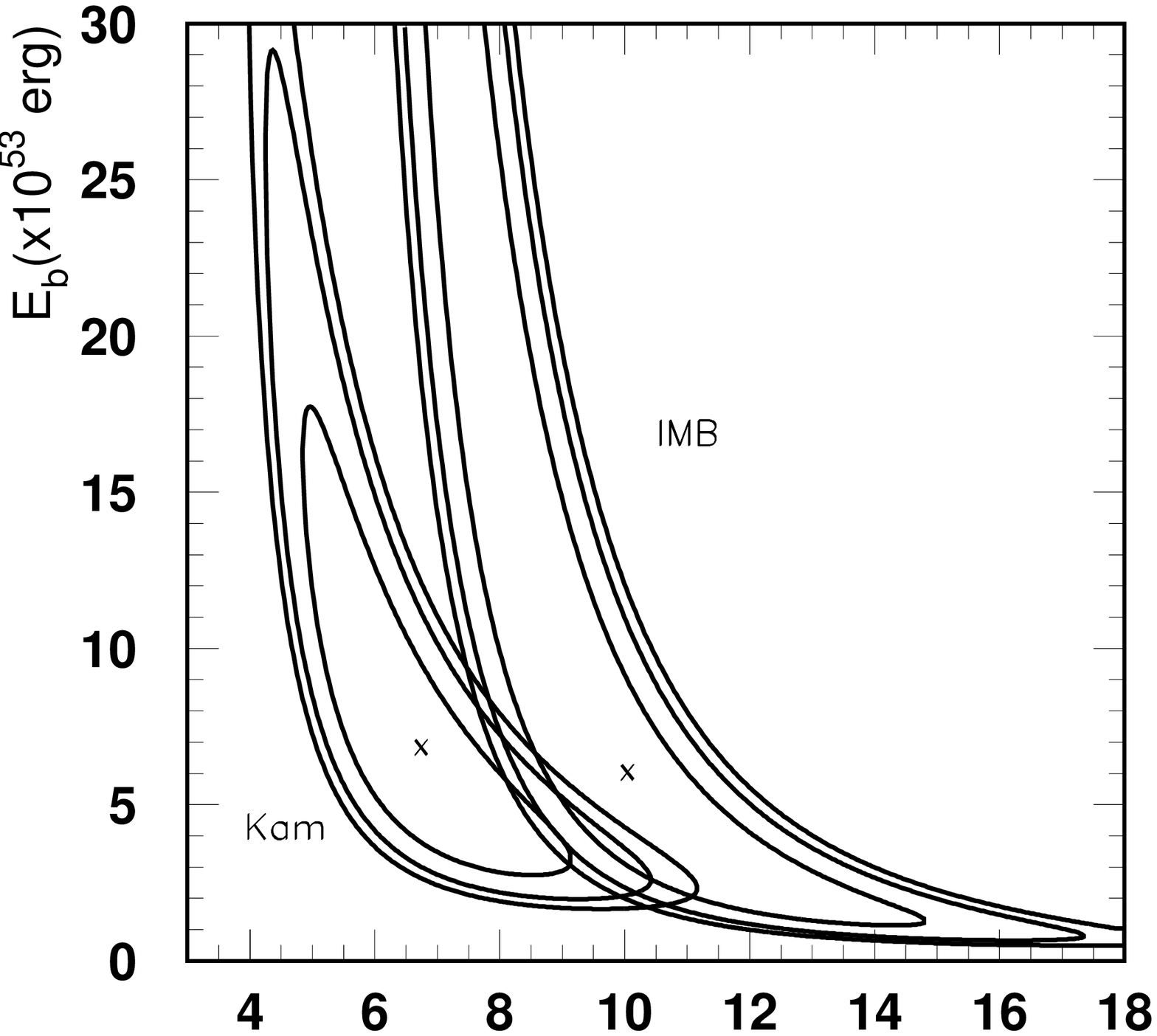,height=8cm,width=12cm,angle=0}
\vspace* {-1.3cm}
           \epsfig{file=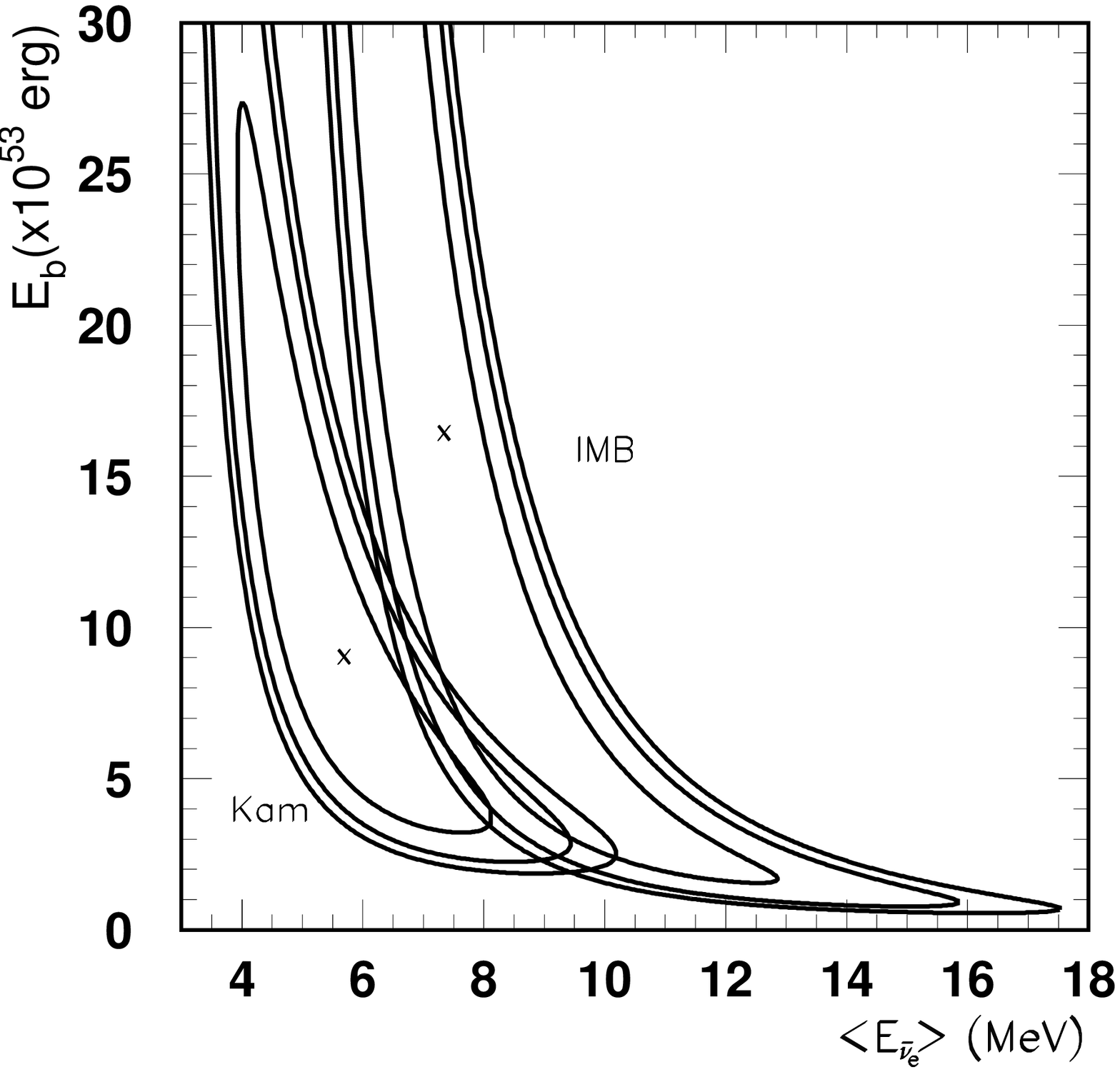,height=8cm,width=12cm,angle=0}
\vspace* {1.2cm}
           \caption{\label{comparison}
             Contours of constant likelihood for the SMA-MSW (top) and
             the LMA-MSW oscillation hypothesis as function of $\Eb$
             and $\Ee$ for $\tau=1.4$ (middle) and $\tau=1.7$
             (bottom). The two experiments are analysed separately.}}

         In the analysis outlined above we simply overlaid the
         theoretical expectations for $\Eb$ and $\Ee$ with the
         confidence regions obtained from the experimental data set.
         In the derivation of these confidence regions we have not
         used any a priori knowledge of $\Eb$ and $\Ee$.  Within the
         Bayesian approach~\cite{Bayesian}, we could combine the
         theoretical prior information $p_{\rm prior}$ from
         Eqs.~(\ref{box1}-\ref{box3}) with the experimental
         information contained in the likelihood function $\L$,
\be
 p_{\rm posterior} (\alpha) \propto \L(\alpha) p_{\rm prior}(\alpha) \,.
\ee
If we are, for instance, absolutely sure that the theoretical range $B$
defined in Eqs.~(\ref{box1}-\ref{box3}) is correct and assume a flat
probability distribution $p_{\rm prior}$ inside this range for $\Eb$
and $\Ee$, we obtain  
\be
 p_{\rm posterior} (\alpha) = \left\{
                              \begin{array}{r}
                              C   \L(\alpha)  \qquad{\rm for}\quad
                              \alpha\in B \phantom{\,.}\\
                              0 \;\quad\qquad{\rm for}\quad
                              \alpha\notin B \,.  \end{array}
                              \right.
\ee
This simple example shows that if one takes serious some theoretical 
predictions $B$, then it is only important how well the different   
oscillations scenarios fit the data inside $B$. On the other hand, the
behaviour of the likelihood functions outside $B$ does not influence 
at all the posterior probability distribution for $\alpha$.
Furthermore, we have noted above that the two fit variables used,
$\Ee$ and $\Eb$, are strongly correlated: Using for instance only the
IMB data set it is impossible to extract any useful information about
$\Eb$. This suggests also to use the a priori knowledge about the
possible range of $\Eb$ and $\Ee$ as input in our analysis.

Another point that can lead to misinterpretations is that the
confidence regions in Figs.~\ref{contour-LMA} and \ref{contour-VO}
were constructed relative to the {\em local\/} minima. Therefore, from
Eq.~(\ref{likelihood.test}), it follows that the large mixing
oscillation solutions are relatively penalized with respect to the
SMA-MSW solution because they have lower $\chi^2$ minima.

\subsection{Testing the different oscillation hypotheses}
\label{hypo}
The ratio of the likelihood functions for different hypotheses
measures the degree to which the experimental data favour one
hypothesis over the other. In order to decide how strong the LMA
solutions are (dis-) favoured against the SMA-MSW solution for a
certain range of astrophysical parameters ($\Eb,\Ee,\tau$) one has
therefore to consider the ratio
\be  \label{R}
 R(\Eb,\Ee,\tau) = \frac{{\cal L}_{\rm LMA}(\Eb,\Ee,\tau)}  
                        {{\cal L}_{\rm SMA}(\Eb,\Ee,\tau)}  \,,
\ee
where we treat $\tau$ as a fixed parameter ($\tau=1.4,~1.7$ or 2
as above). Then the probability $\beta_{\rm LMA}$ that the LMA
and not the SMA hypothesis is compatible with the data for certain 
astrophysical parameters is given for $\beta\to 1$ or $\beta\to 0$ 
approximately by the usual $\Delta\chi^2_{\beta;2} = - 2\ln(R)$. 
If both hypotheses have roughly the same probability, 
i.e. $\Delta\chi^2_{\beta;2}\ap 0$, the equation 
$\beta_{\rm LMA}(\chi^2_{\rm LMA})+\beta_{\rm SMA}(\chi^2_{\rm LMA}+c)=1$
has to be solved for a given contour $\Delta\chi^2=c$,
cf. Table~\ref{chi-prob}. 
\begin{table}
\begin{center}
\begin{tabular}{|c|c|c|c|c|c|} 
\hline 
$ \Delta\chi^2_{\beta;2}=-2\ln(R)$ & 
0  & 2 & 4 & 6 & 10 \\ \hline
$\beta_{\rm LMA}=1-\beta_{\rm SMA}$ &
50\% & 27\% & 12\% & 5\% & 0.7\% \\ \hline             
\end{tabular}
\end{center}
\caption{\label{chi-prob}
Probability $\beta$ that one of the two hypotheses describes the data.}
\end{table}

In Eq.~(\ref{R}), we compare the likelihood of different oscillation
hypotheses for the same astrophysical parameters. Therefore, we can
now answer questions like ``how large is the maximal probability that
the LMA-MSW solution is compatible with the neutrino signal, assuming
that the predictions for $\Eb$, $\Ee$ and $\Eh$ of a certain
simulation are correct?'' 

In Figs.~\ref{lma-sma} and \ref{vac-sma} we show the likelihood ratios
$\ln(R)$ of the SMA-MSW hypothesis compared to LMA-MSW and the VO,
respectively. Let us discuss first the LMA-MSW solution. Restricting
the possible $\Eb,\Ee,\tau$ values to the theoretically favoured
region~(\ref{box1}-\ref{box3}), the SMA-MSW solution is always
favoured. In particular, the LMA-MSW hypothesis can be clearly
excluded for large values of $\Eb$ together with large $\tau$. The
preference for the SMA-MSW solution is however weak for low values of
$\Eb$, $\Ee$ and $\tau$ and looses any statistical significance for
$\Eb=1.5\times 10^{53}$~erg, $\Ee=14$~MeV and $\tau=1.4$. For $\Ee\ap
12$~MeV, e.g. neutrino energies predicted by recent simulations, there
are even points in the $\Ee$-$\Eb$ plane for which LMA-MSW is {\em
  favoured\/} compared to the SMA-MSW solution. Note also that this
region of relatively small $\Ee$ is favoured by the neutrino signal of
SN~1987A for {\em all\/} three oscillation solutions.

\FIGURE{\vspace*{-1.3cm}
        \epsfig{file=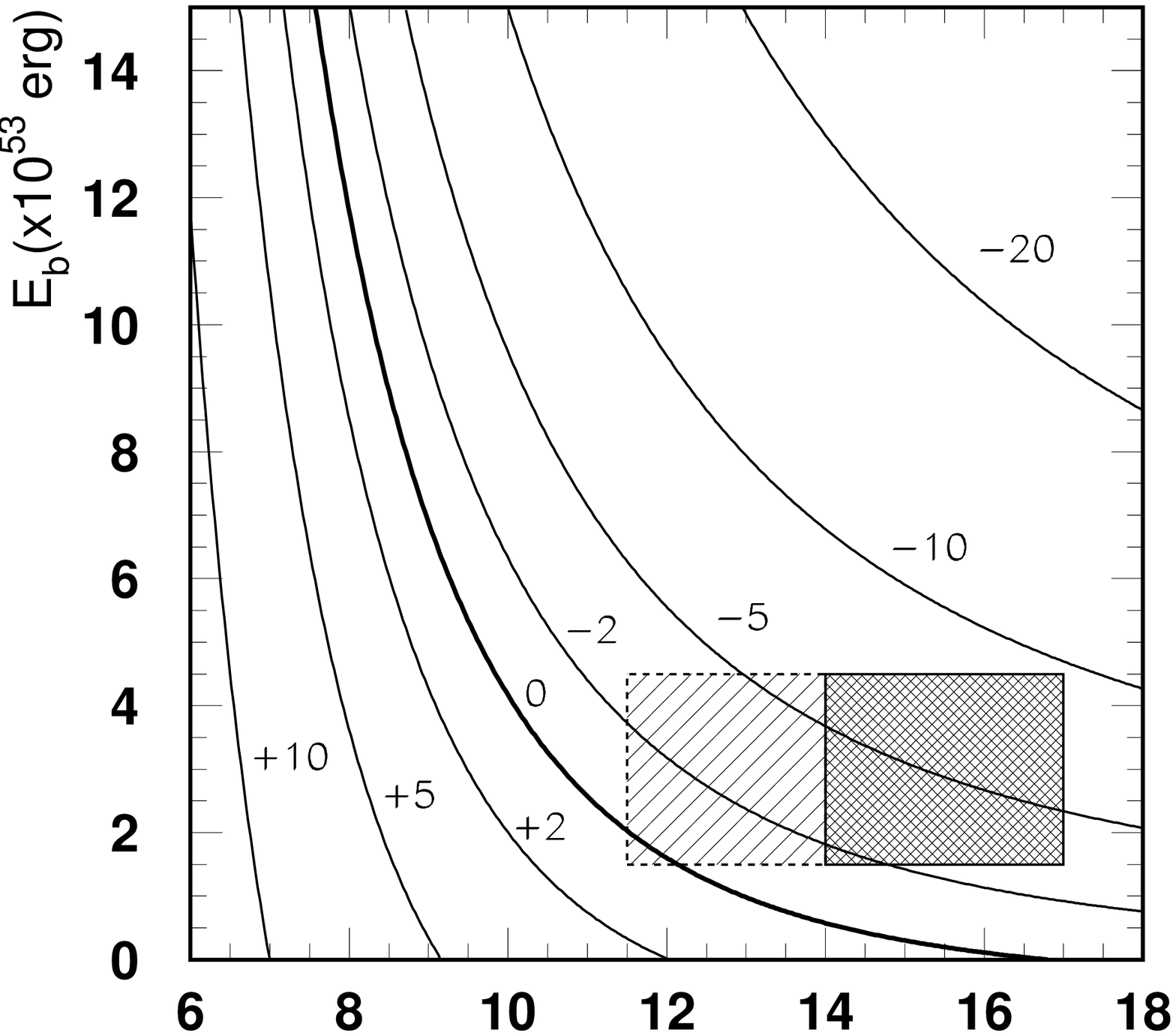,height=8.2cm,width=12.0cm}  
        \vspace*{-1.3cm}
        \epsfig{file=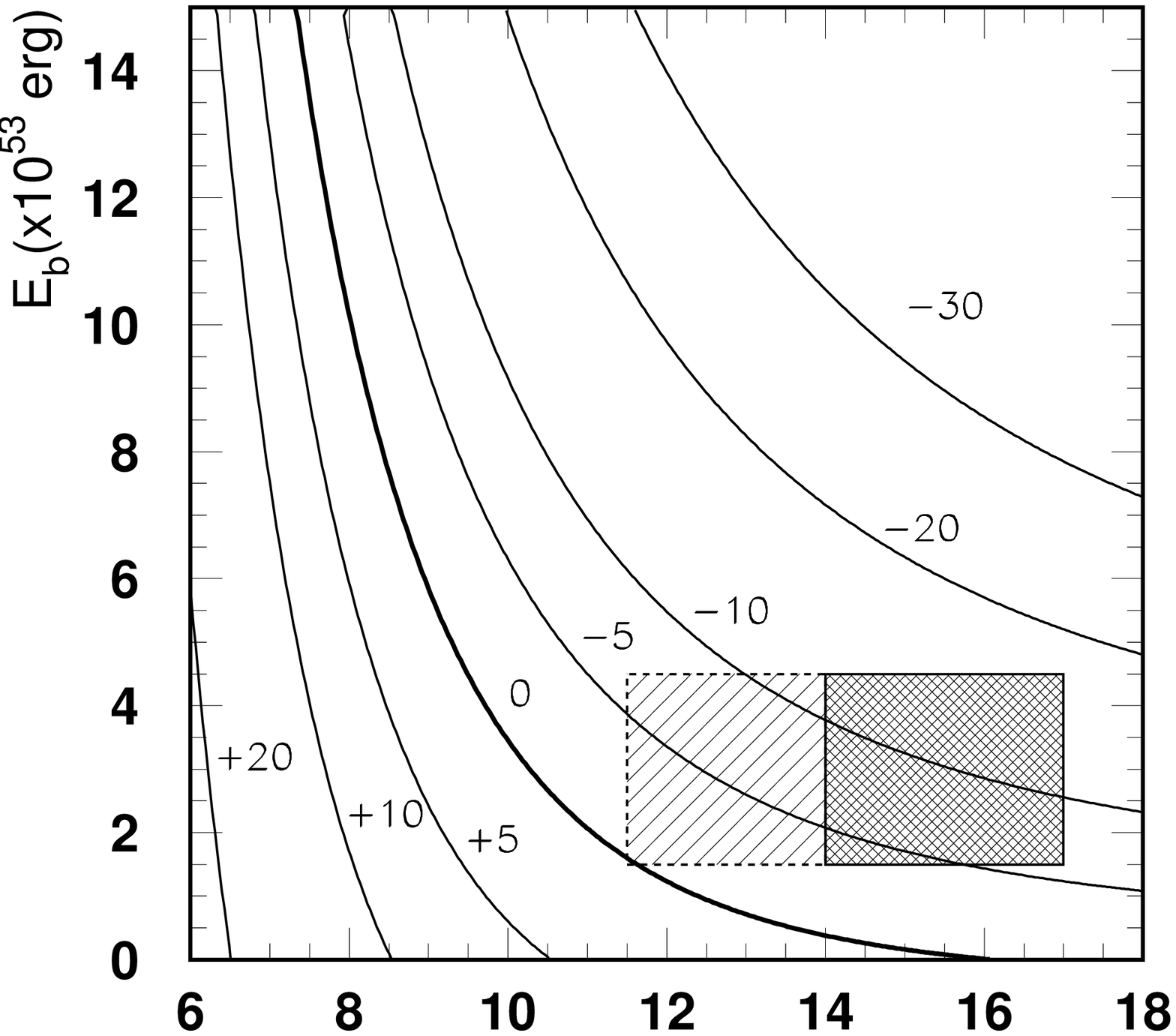,height=8.2cm,width=12.0cm} 
        \vspace*{-1.3cm}
        \epsfig{file=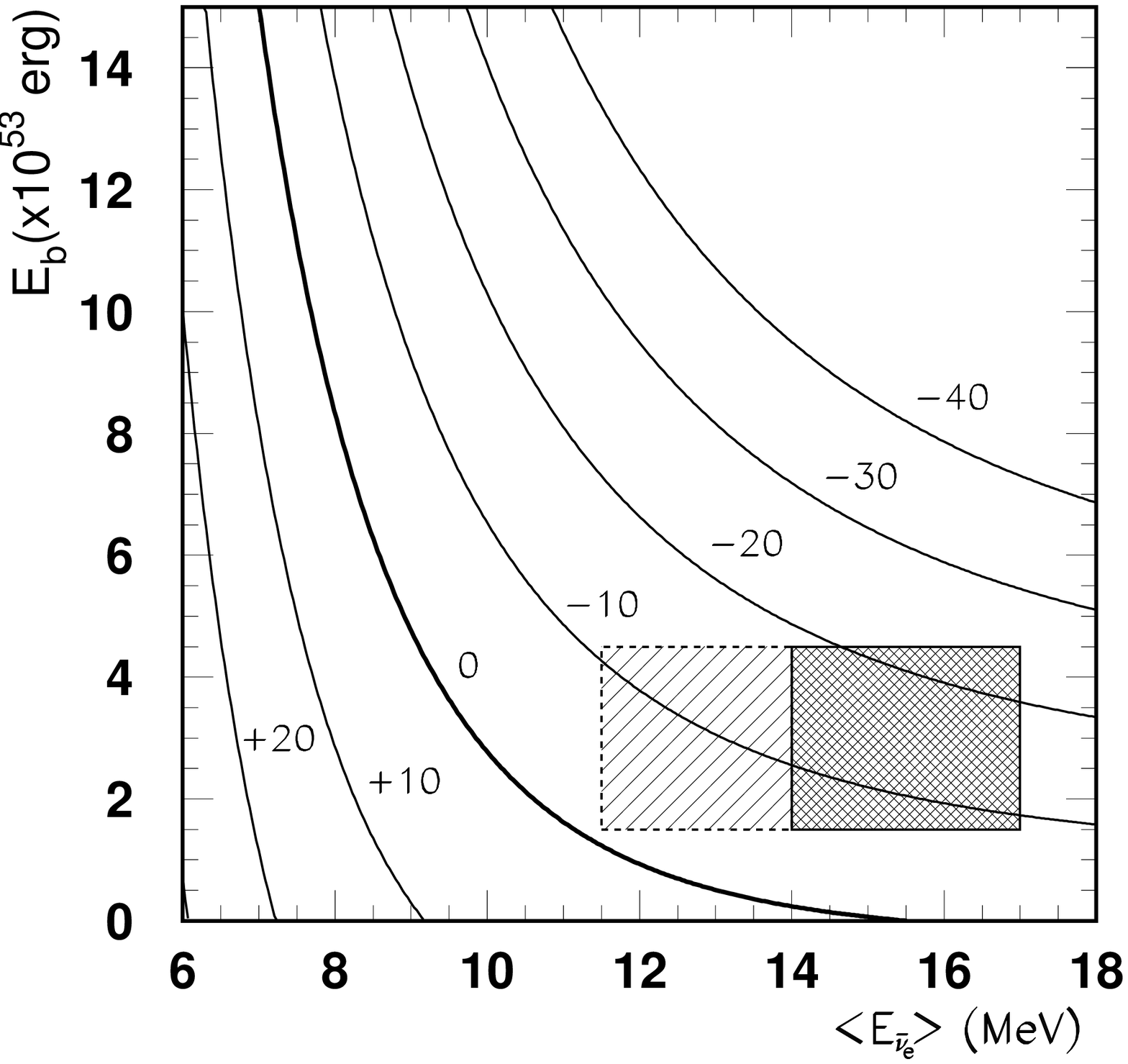,height=8.2cm,width=12.0cm}  
        \vspace*{1.2cm}
        \caption{\label{lma-sma}
        Likelihood ratio $\ln(R)$ of the LMA and SMA oscillation hypothesis
        as function of $\Eb$ and $\Ee$ for $\tau=1.4$ (top),
        $\tau=1.7$ (middle) and $\tau=2$ (bottom).}}

    The case is different for the VO solution. Within the region
    (\ref{box1}-\ref{box3}), the likelihood ratio is $-150\lsim
    2\ln(R)\lsim -10$. In particular, large values of $\tau$ and $\Eb$
    are clearly excluded by the data. Only if one allows for rather
    small neutrino energies, $\Ee\lsim 12$~MeV and $\tau=1.4$, some
    parameter space remains for which the VO solutions is not clearly
    disfavoured by the data.

\FIGURE{\vspace*{-1.3cm}
        \epsfig{file=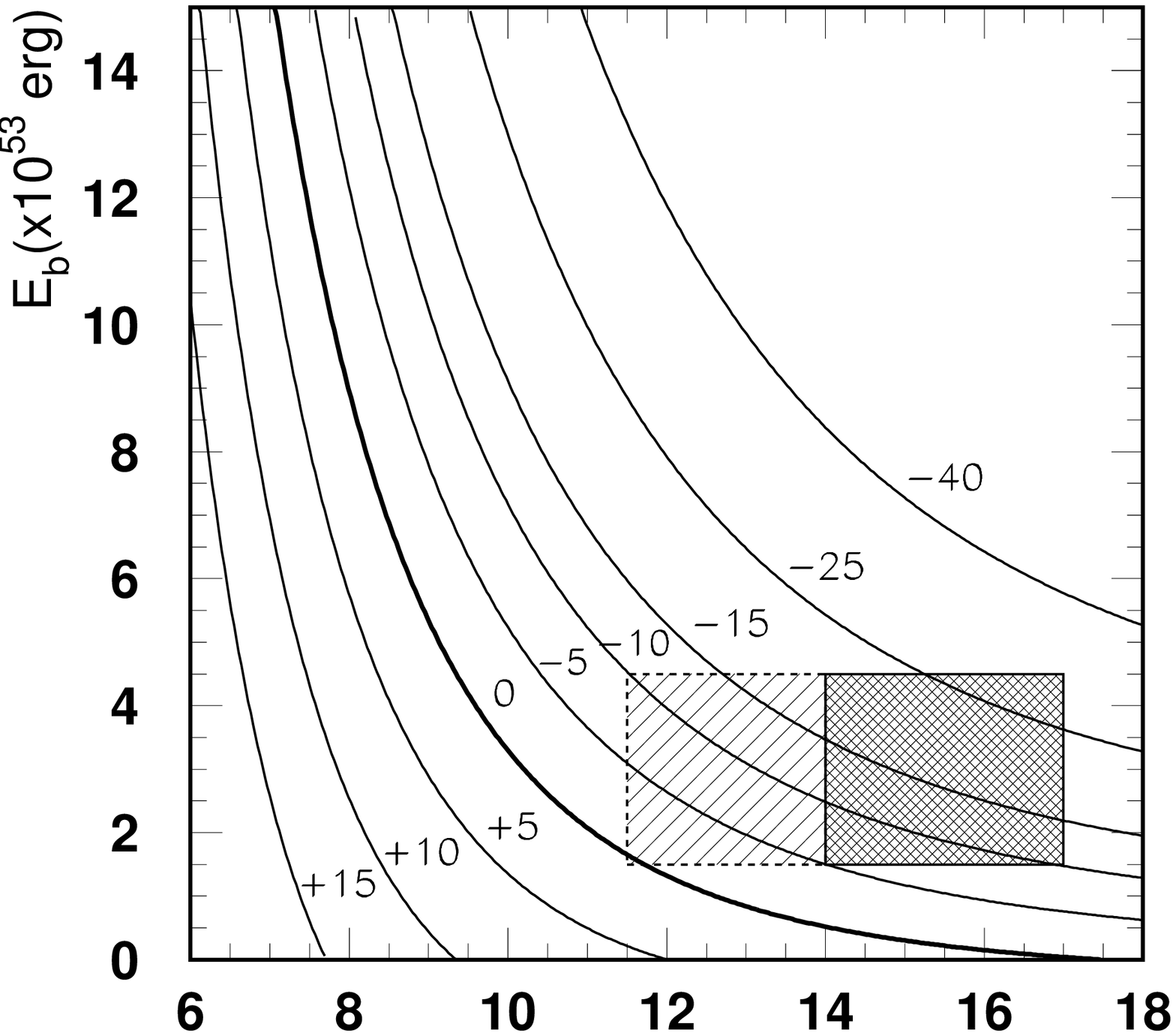,height=8.2cm,width=12.0cm}  
        \vspace*{-1.3cm}
        \epsfig{file=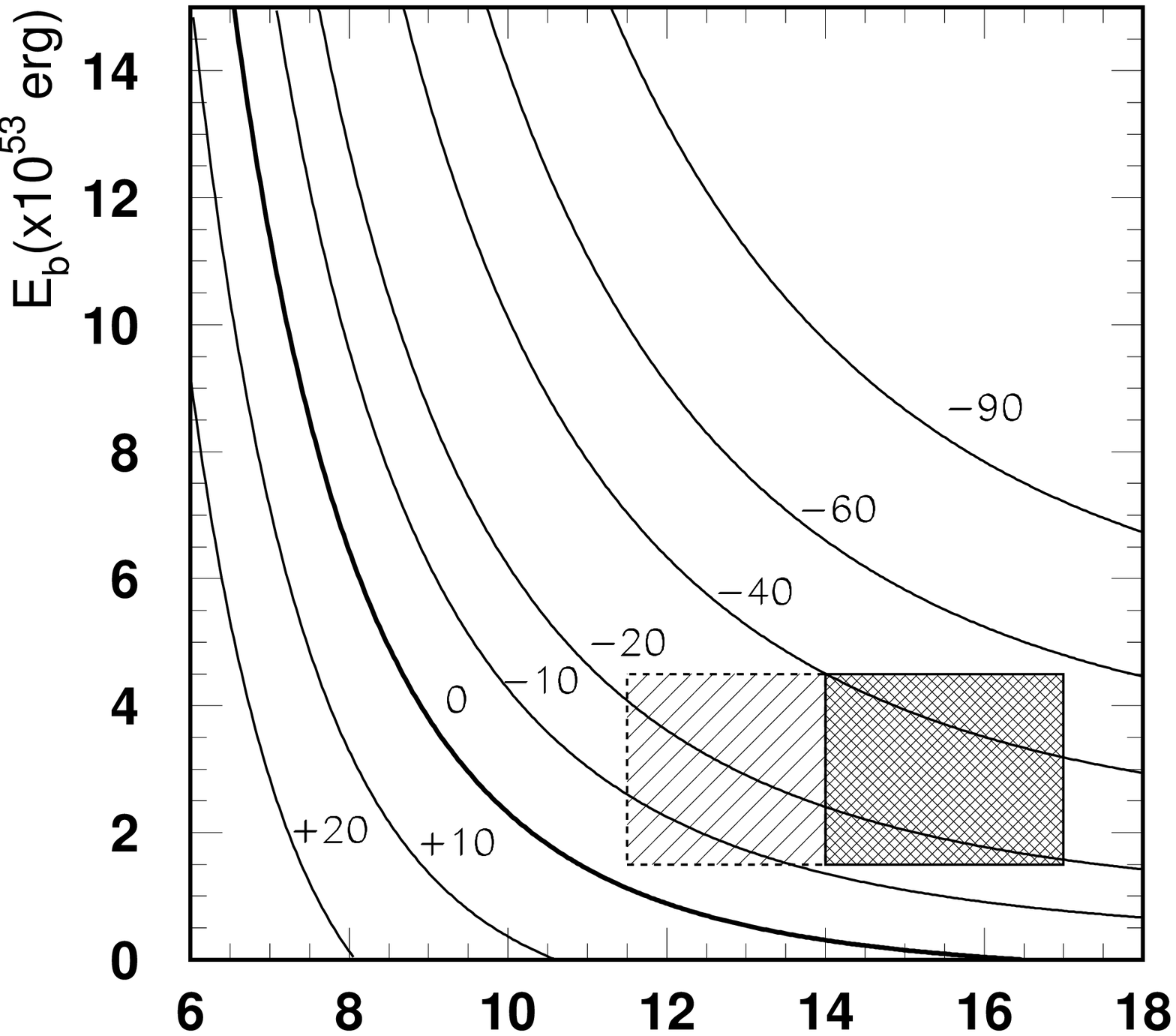,height=8.2cm,width=12.0cm}  
        \vspace*{-1.3cm}
        \epsfig{file=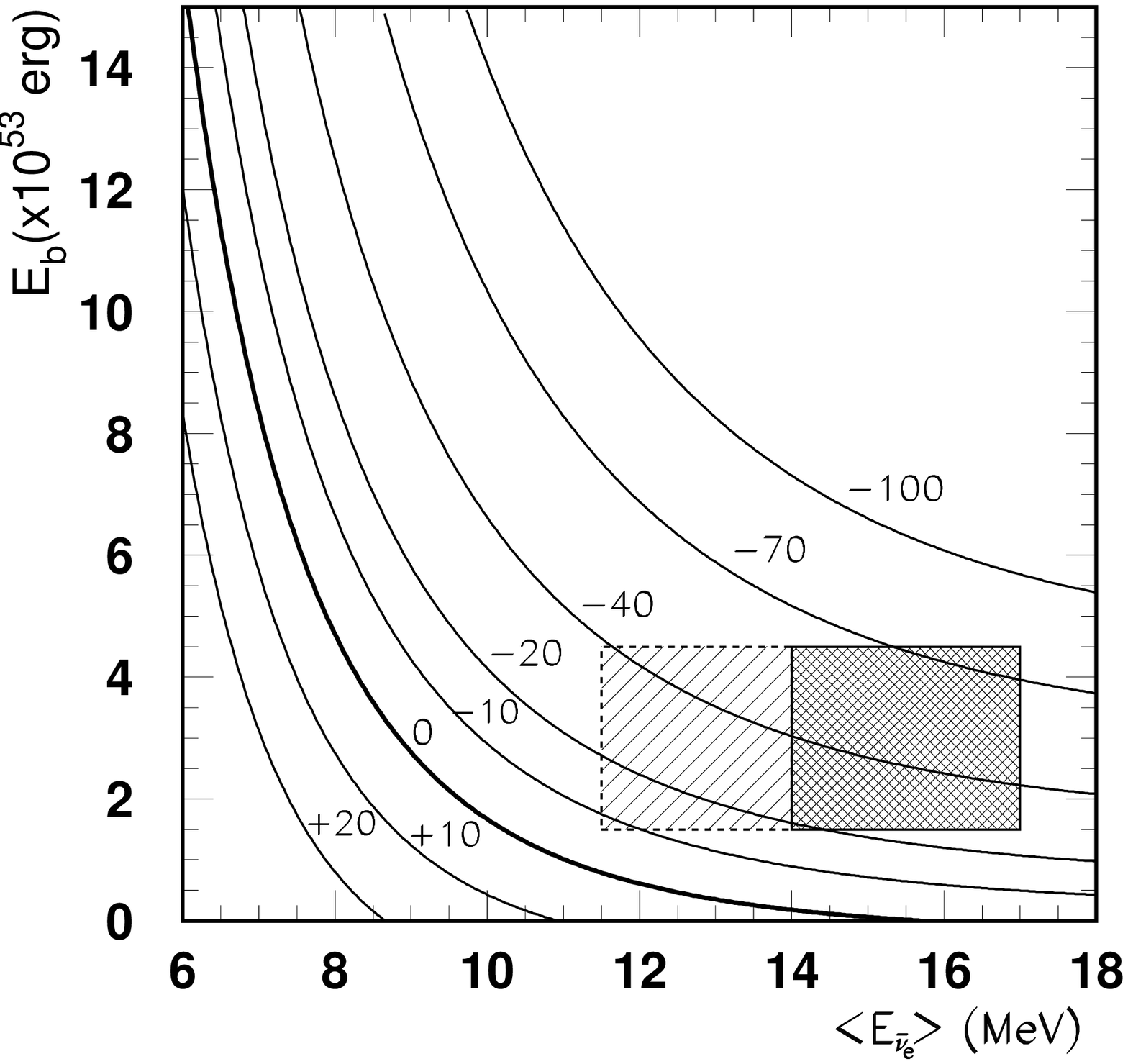,height=8.2cm,width=12.0cm}  
        \vspace*{1.4cm}
        \caption{\label{vac-sma}
        Likelihood ratio $\ln(R)$ of the VO and SMA oscillation
        hypothesis as  function of $\Eb$ and $\Ee$ for $\tau=1.4$
        (top), $\tau=1.7$ (middle) and $\tau=2$ (bottom).}}

    Finally we want to discuss the case of the LOW solution to the
    solar neutrino problem. As has been shown in Ref.~\cite{global00}
    this solution also provides a good global fit of the solar
    neutrino data and extends continuously from the ``normal''
    $\theta<\pi/4$ (light side) case into the region with
    $\theta>\pi/4$, the so--called dark-side. 
    In order to determine the impact of $\bar\nu_e \leftrightarrow
    \bar\nu_{\mu,\tau}$ neutrino oscillations in the LOW region on the
    observed $\bar\nu_e$ signal of supernova SN 1987A we have repeated
    a likelihood analysis for $\theta$ and $\dm$ for this case,
    choosing for $\Eb$, $\Ee$ and $\tau$ some representative values.
    In Figs.~\ref{dark_12} and \ref{dark_14} we show the likelihood
    ratio $\ln(R)$ as function of $\tan^2\theta$ and $\dm$ relative to
    the SMA hypothesis for two values of the average $\bar\nu_e$
    energy, $\Ee=12$~MeV and $\Ee=14$~MeV, respectively.  
    The contours of constant likelihood shown correspond to
    $\ln(R)=-1,-2,-3,-5,-10,-15,-20,-30$, if not otherwise indicated. 
    We find that the contours are symmetric with respect to
    $\theta=\pi/4$ below $10^{-7}$-$10^{-8}$~eV$^2$.
    Thus matter effects do not play a role in neither the SN envelope
    nor the Earth for most of the LOW region.
    Therefore, the dark side is as compatible with the SN~1987A
    neutrino signal as the light side.
    The absence of Earth matter effects below $10^{-5}$~eV$^2$ also
    shows that regeneration effects are negligible for the LOW and VO
    solutions, and ``explains'' why their status is worse than that of
    the LMA solution from the point of view of the present supernova
    analysis.
    
    In Figs.~\ref{dark_12} and \ref{dark_14} we present also the 90,
    95 and 99\% C.L.
    contours of the other possible solutions to the solar neutrino
    problem (from Ref.~\cite{global00}).
    These figures summarize therefore our
    findings above. The Earth matter effect regenerate partially the
    $\bar\nu_e$ fluence for $\dm\sim 10^{-5}$~eV$^2$ and thereby
    increases the likelihood of the LMA-MSW solution when compared to
    the LOW and VO solutions.
    The LMA-MSW solution has a large probability to be compatible with
    the SN1987 data for a considerable part of the space of
    astrophysical parameters, while the LOW and VO solutions are
    generally strongly disfavoured by the data, except for the extreme
    case low $\Eb$, $\Ee$ and $\tau$.
 
\FIGURE{
        \epsfig{file=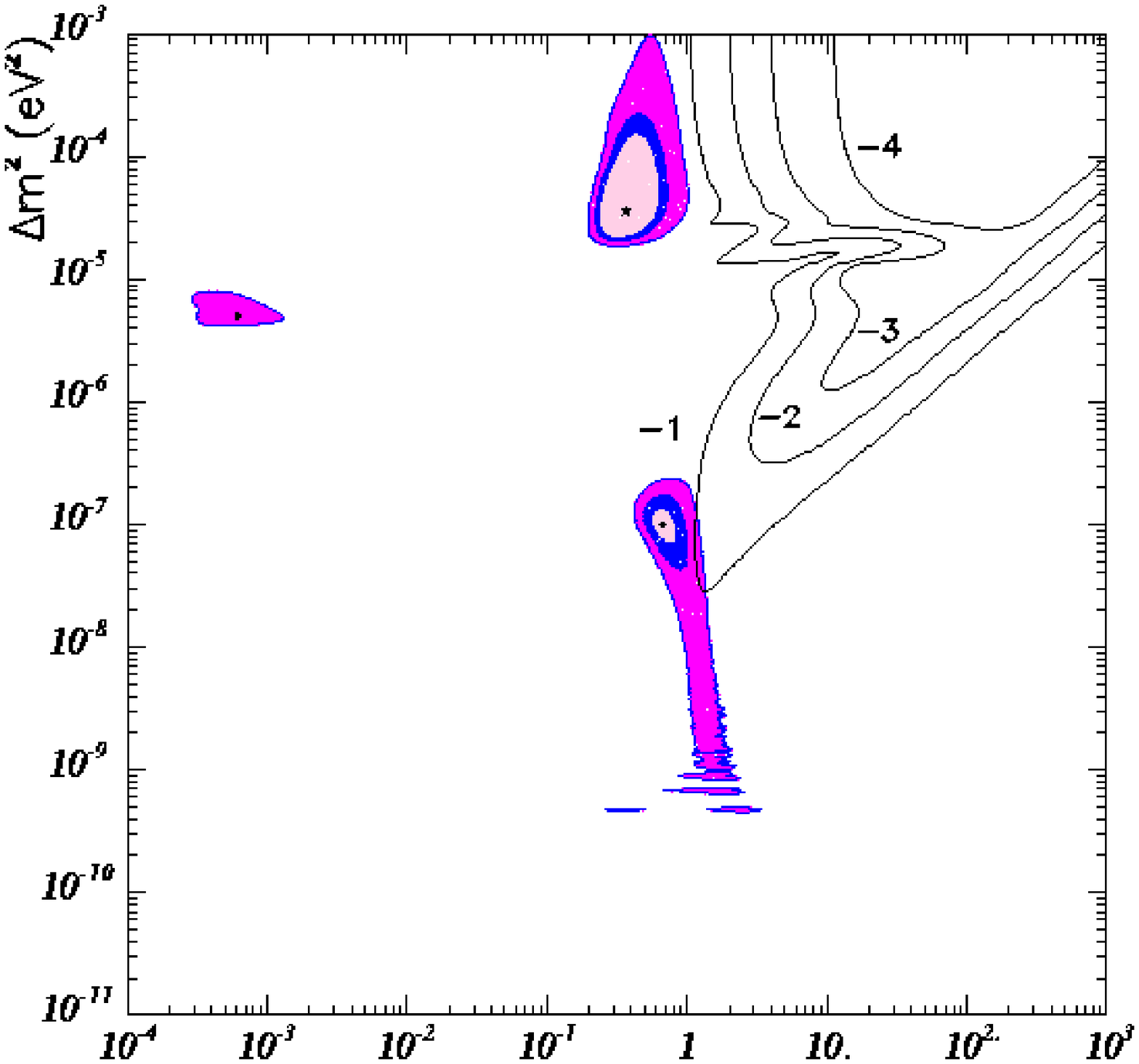,height=7.3cm,width=12.0cm}  
        \epsfig{file=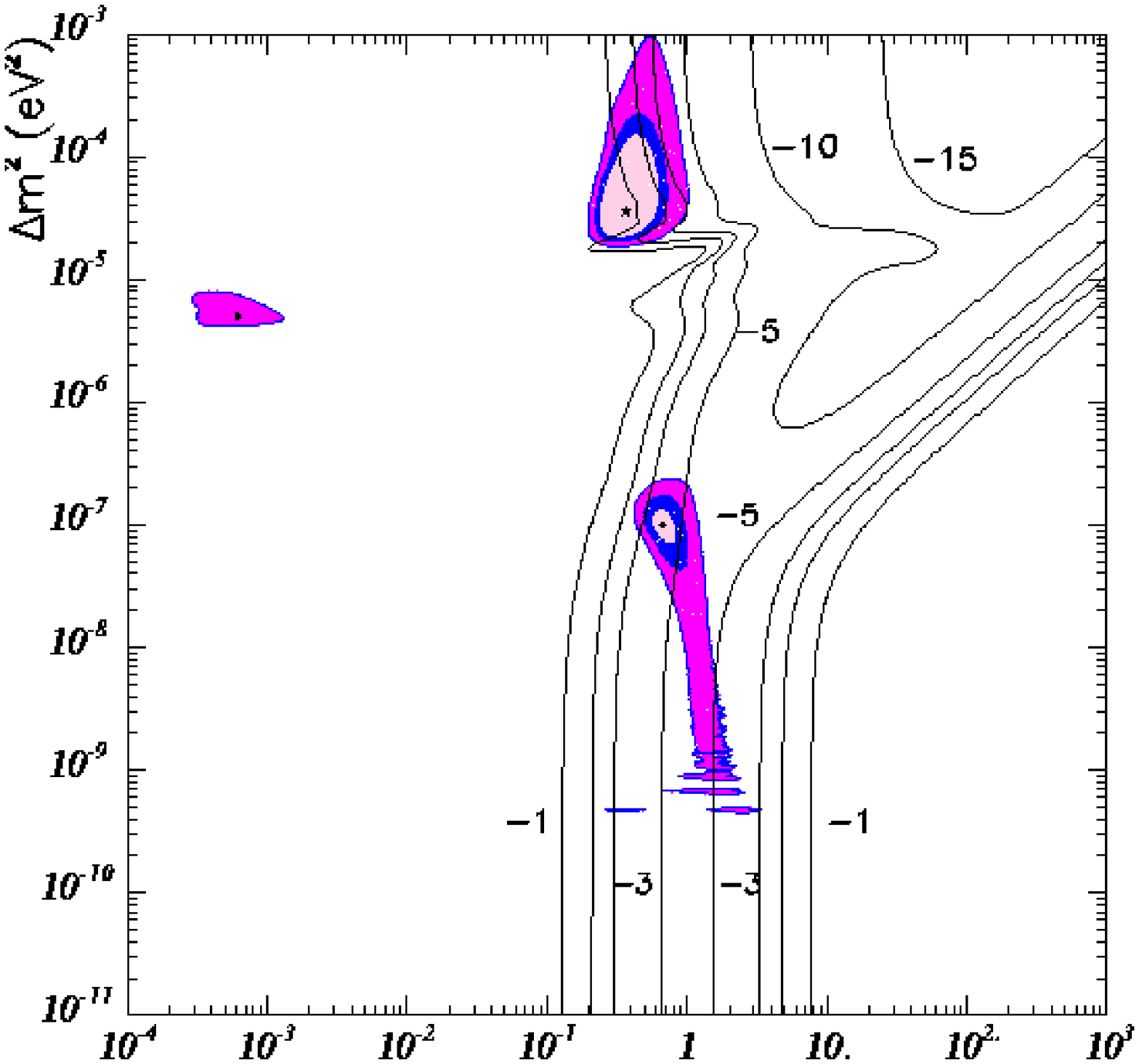,height=7.3cm,width=12.0cm} 
        \epsfig{file=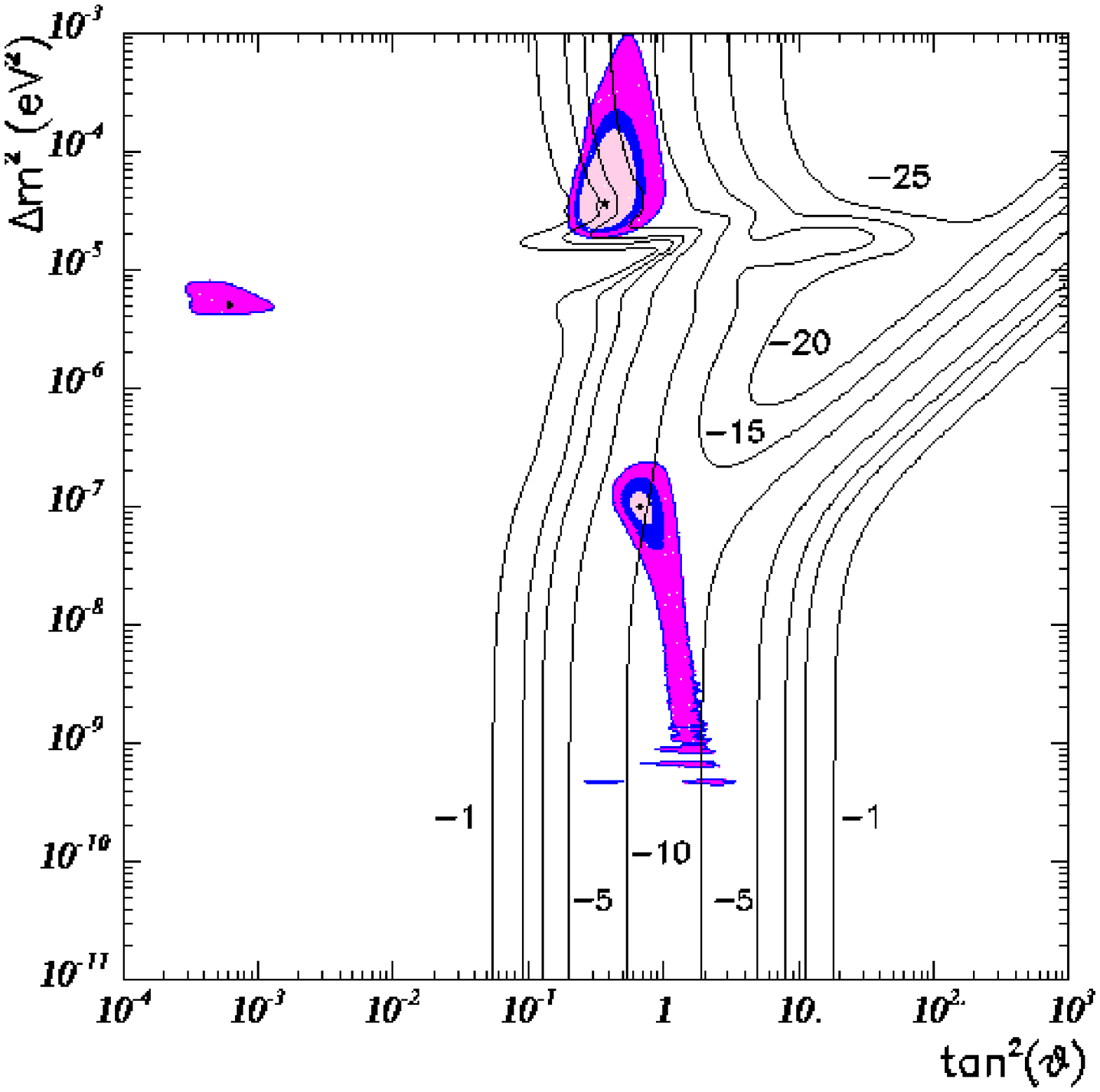,height=7.3cm,width=12.0cm}  
        \caption{\label{dark_12}
          Likelihood ratio $\ln(R)$ relative to the SMA-MSW
          hypothesis, as function of $\tan^2\theta$ and $\dm$/eV$^2$
          for $\tau=1.4$ (top), $\tau=1.7$ (middle) and $\tau=2$
          (bottom) together with the 90, 95 and 99\% CL contours for
          the different solutions to the solar neutrino problem from
          Ref.~\cite{global00}. All figures for $\Eb=1.5 \times
          10^{53}$~erg and $\Ee=12$~MeV.}}
 
\FIGURE{
        \epsfig{file=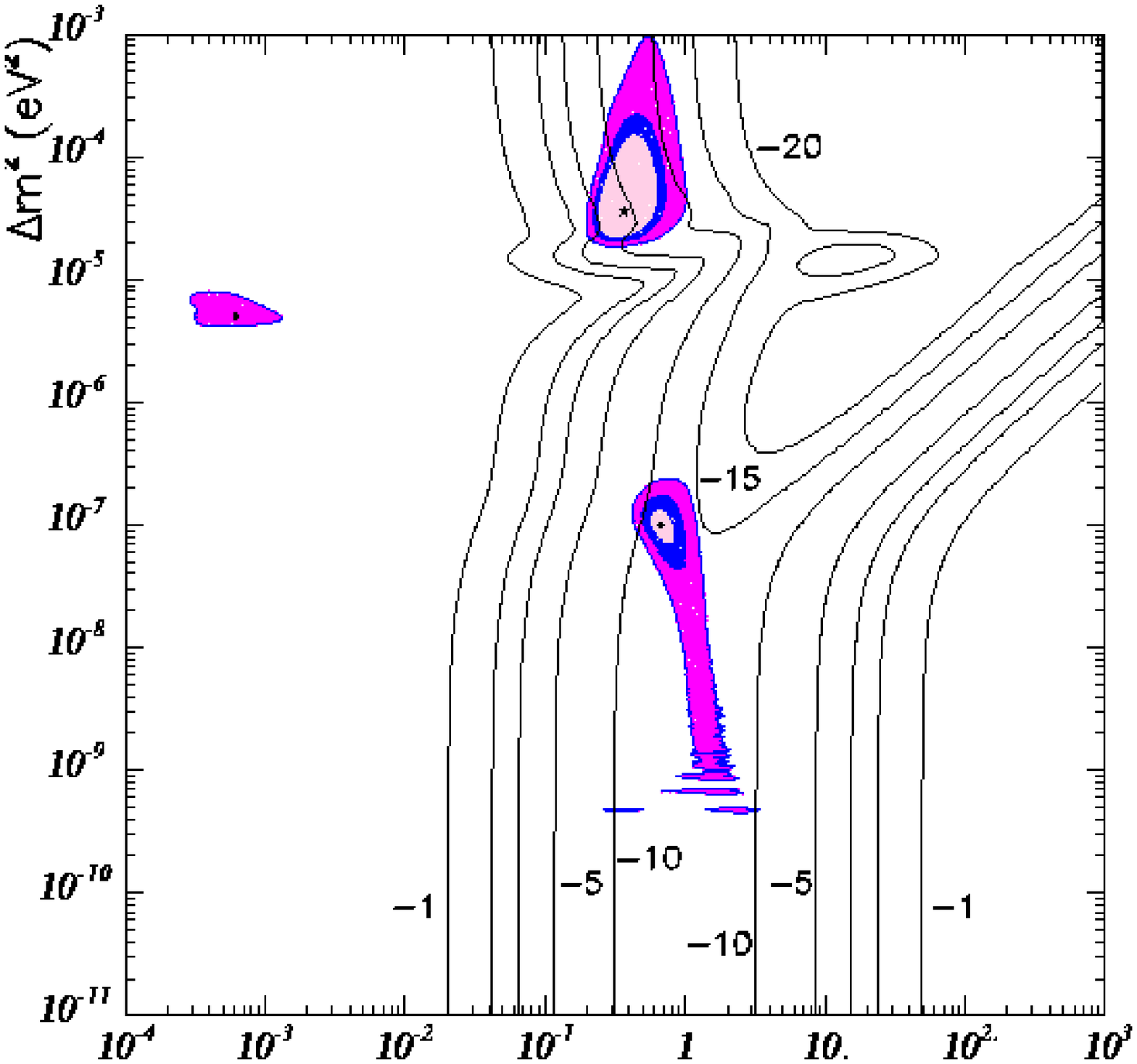,height=7.3cm,width=12.0cm}  
        \epsfig{file=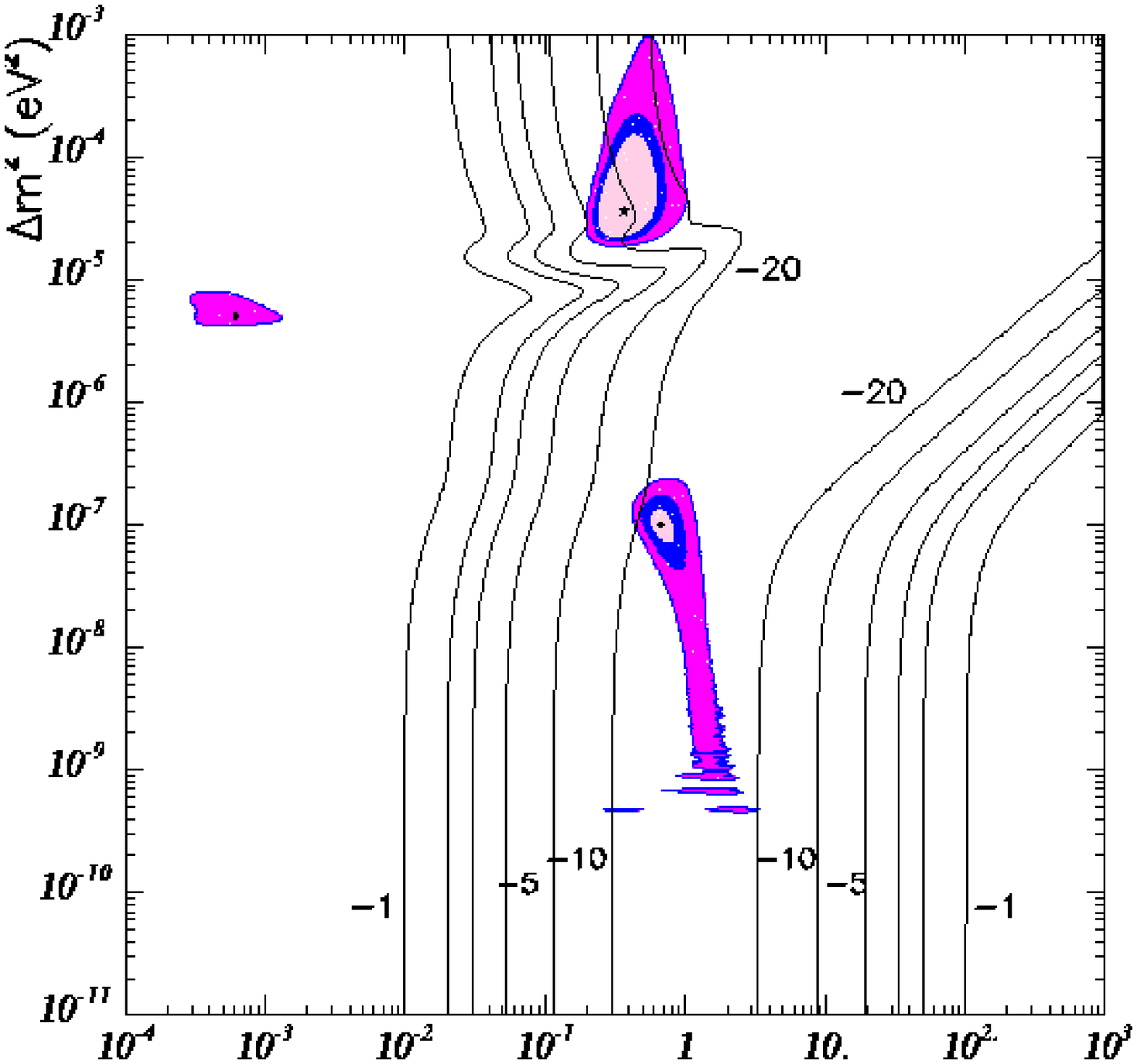,height=7.3cm,width=12.0cm} 
        \epsfig{file=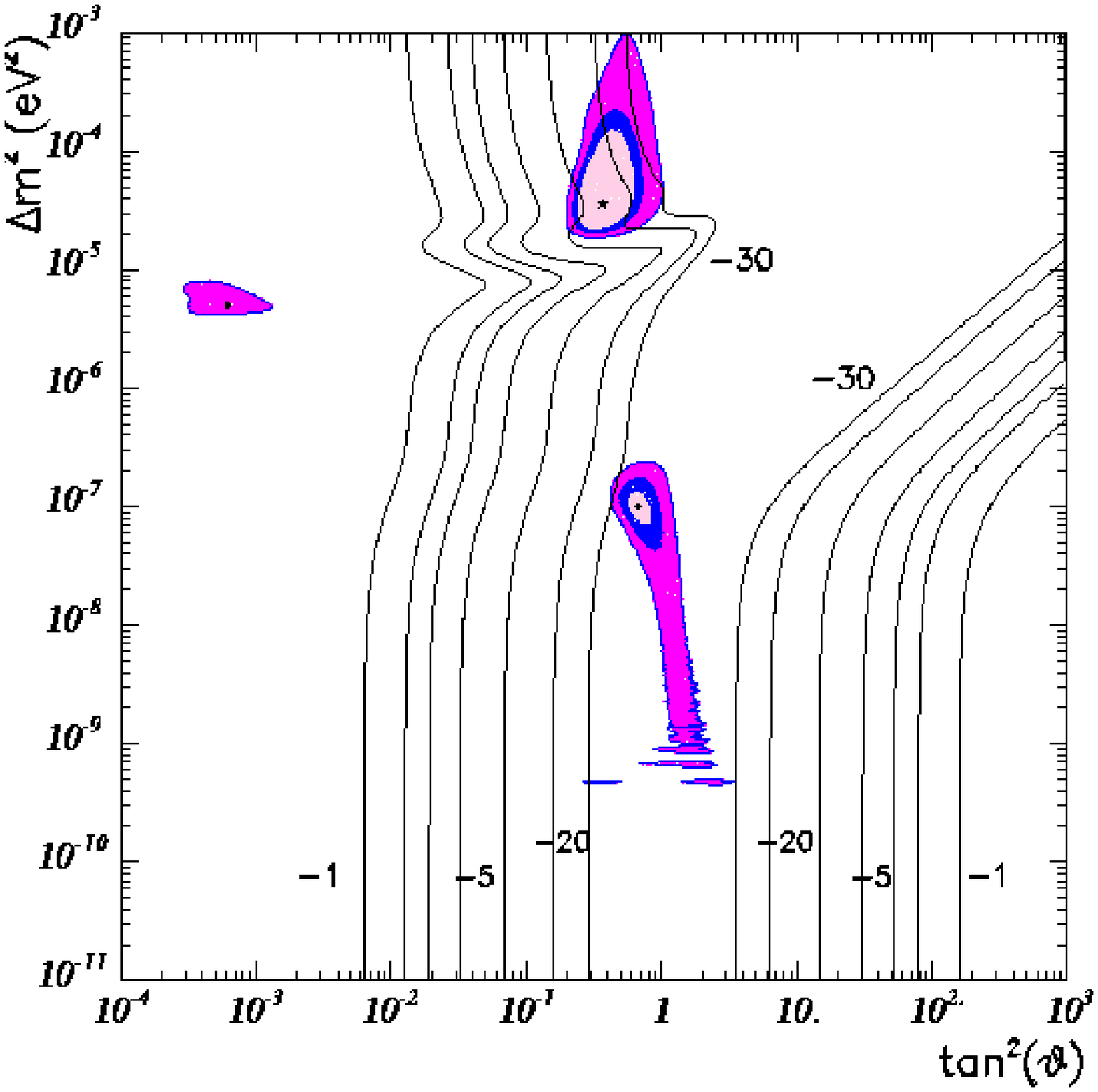,height=7.3cm,width=12.0cm}  
        \caption{\label{dark_14}
          Likelihood ratio $\ln(R)$ relative to the SMA-MSW
          hypothesis, as function of $\tan^2\theta$ and $\dm$/eV$^2$
          for $\tau=1.4$ (top), $\tau=1.7$ (middle) and $\tau=2$
          (bottom) together with the 90, 95 and 99\% CL contours for
          the different solutions to the solar neutrino problem from
          Ref.~\cite{global00}. All figures for $\Eb=3\times
          10^{53}$~erg and $\Ee=14$~MeV.}}

\section{Kolmogorov-Smirnov test}
We use the one-dimensional Kolmogorov-Smirnov (KS) test as described
e.g. in Ref.~\cite{nr} to estimate the probability that the observed
data points $E_i$ agree with the spectral shape  predicted by a given
distribution $n_{\bar\nu_e}(E)$. The one-dimensional form of the KS
test we apply has the disadvantage of being not sensitive to $\Eb$ viz.
the amplitude of the neutrino signal. This disadvantage is however
compensated by its property to be distribution-free that makes the
one-dimensional KS test attractive. In
Fig.~\ref{KS1_SMA}, we show the probability that the signal observed
in the two detectors is consistent with the SMA-MSW hypothesis, as
function of $\Ee$.  Performing this test for the two experiments
separately, we find a small region of overlap at $\Ee\sim 11$~MeV
where this hypothesis has a probability of more than 10\% in both
experiments.  The combined data set has a probability higher than 10\%
in the broad range $\Ee= 9~$MeV -- 14~MeV, with a maximal probability
of 46\%.  Since the maximum occurs at the same energies as the
best-fit values found in the maximum-likelihood analysis, we conclude
that the essential information is contained in the form of the
spectra, not in the absolute number of events.

In Fig.~\ref{KS1_LMA}, we show the same test for the LMA-MSW
hypothesis. Its probability as function of $\Ee$ has the same shape as
in the case of SMA-MSW but is shifted to lower energies by 0.7~MeV
($\tau=1.4$), 1.7~MeV ($\tau=1.7$) and 2.8~MeV ($\tau=2$),
respectively. Consequently, an exclusion of the LMA-MSW solution
because of its preference of low $\Ee$ requires a correspondingly
precise knowledge of the average neutrino energies emitted by the SN.
The maximum of the probability as function of $\Ee$ increases from
50\% ($\tau=1.4$) to 65\% ($\tau=1.7$) and 80\% ($\tau=2$).  Moreover,
we find that the LMA-MSW hypothesis is consistent with the data at the
5\% ($\tau=1.4$), 2\% level ($\tau=1.7$) and 1\% level ($\tau=2$) for
$\Ee=14$~MeV, while the compatibility increases to 25\% ($\tau=1.4$),
16\% level ($\tau=1.7$) and 5\% level ($\tau=2$) at the lower end of
predicted neutrino energies, $\Ee=12$~MeV.

Finally, we present in Fig.~\ref{KS1_VO} the results of the KS test
for VO. Now the maxima of the probability as function of $\Ee$
compared to the SMA-MSW solution are shifted by 2.2~MeV ($\tau=1.4$),
3.7~MeV ($\tau=1.7$) and 5~MeV ($\tau=2$), respectively. The maximum
of the probability increases from 59\% ($\tau=1.4$) to 68\%
($\tau=1.7$) and 70\% ($\tau=2$). Thus the most probable value of
$\Ee$ for $\tau=2$ is reduced by a factor two  compared to SMA-MSW and
also the cases $\tau=1.7$ and $\tau=1.4$ are highly incompatible with the
range of neutrino temperatures predicted by simulations. On the other hand,
VO together with rather low neutrino temperatures, $\Ee\ap 12$~MeV and
$\tau=1.4$ are at the $2\sigma$ level compatible with the spectral
shape of SN~1987A signal.

\FIGURE{\epsfig{file=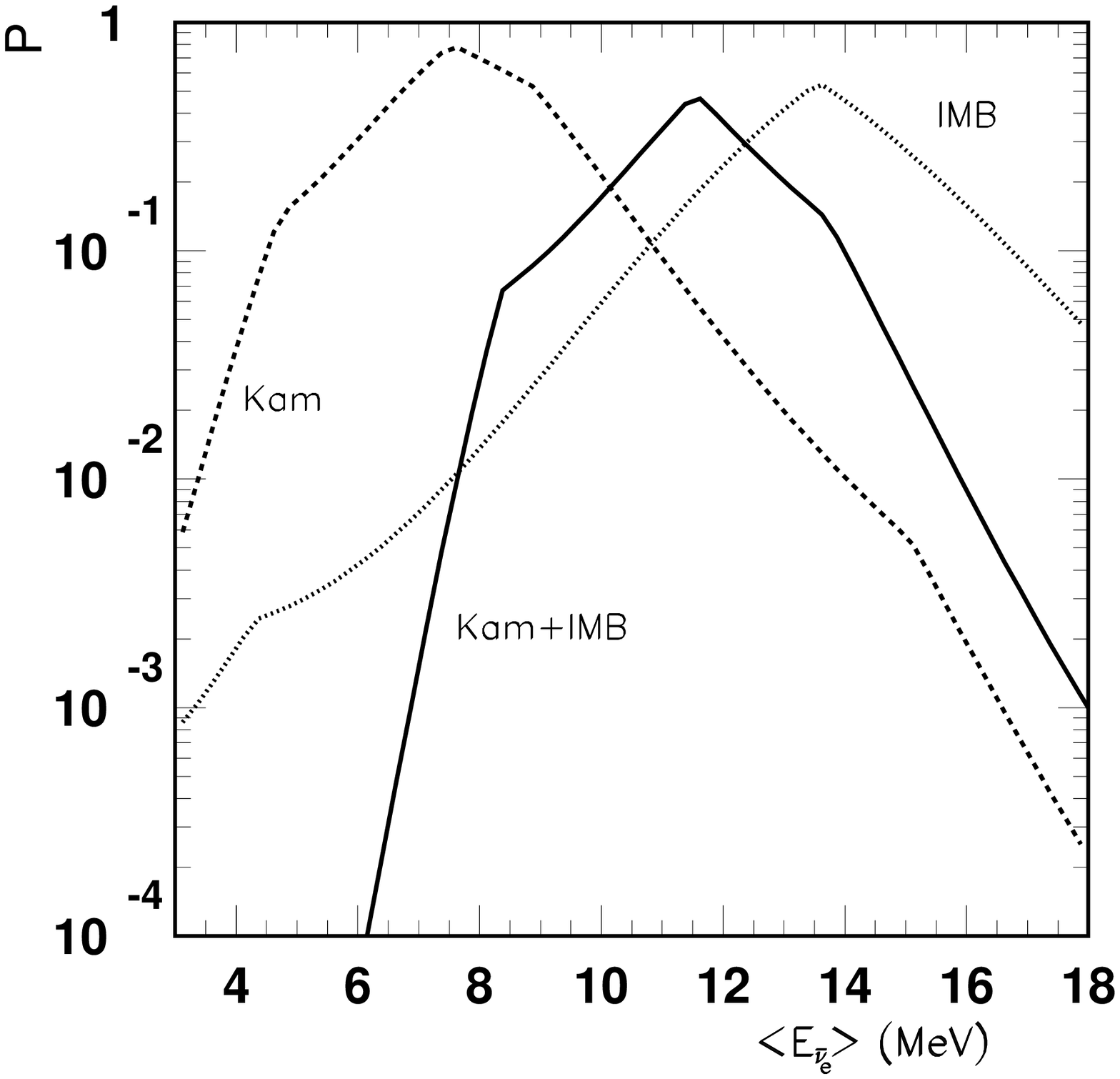,height=8.5cm,width=12.0cm}
        \caption{\label{KS1_SMA}
        Probability $p$ as function of $\Ee$ given by the KS test
        that the observed data by
        Kam, IMB and the combined data set are consistent with
        the SMA-MSW oscillation hypothesis.}}

\FIGURE{\epsfig{file=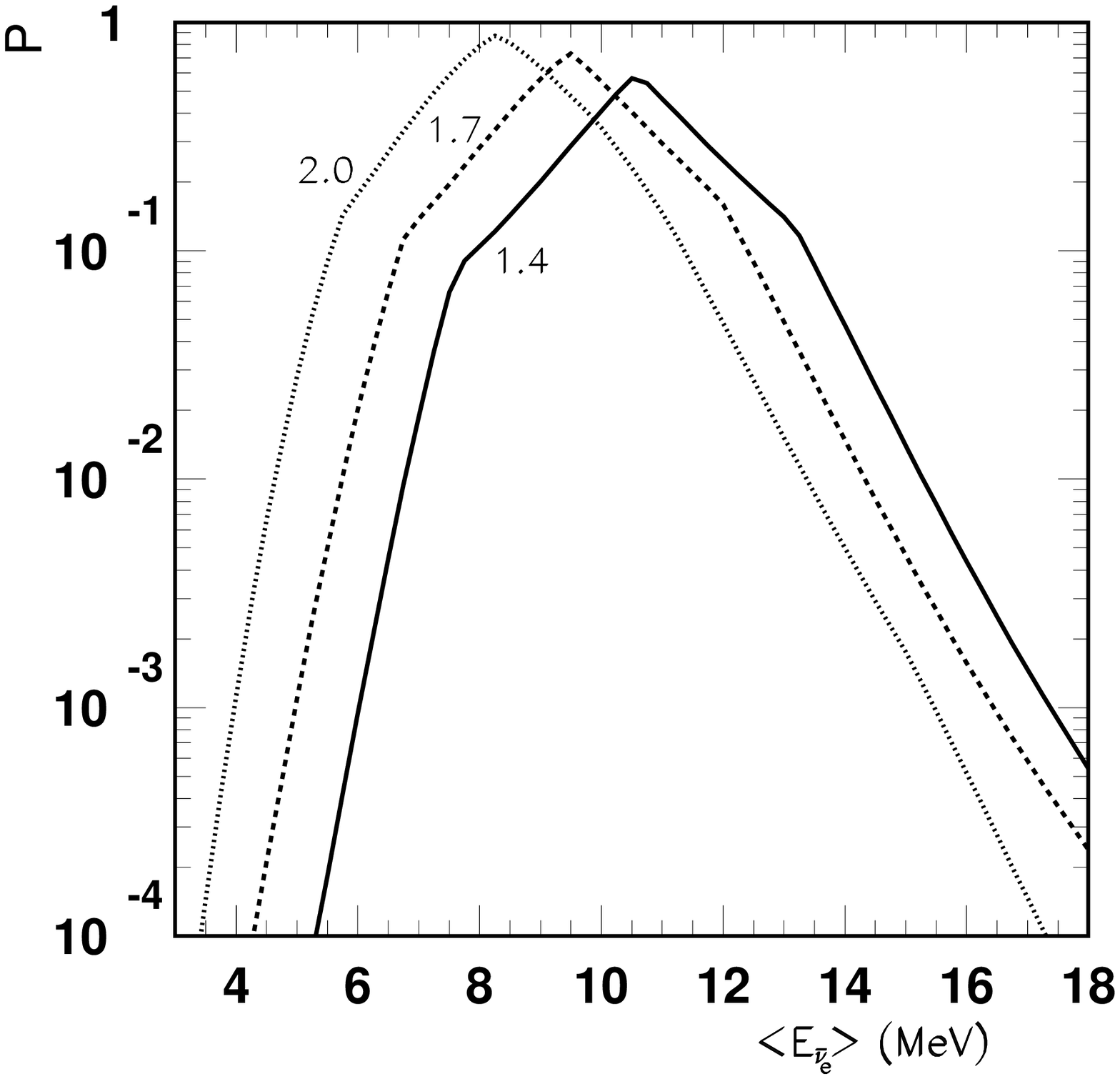,height=8.5cm,width=12.0cm}
        \caption{\label{KS1_LMA}
        Probability $p$ as function of $\Ee$ given by the KS test
        that the combined data set is consistent with
        the LMA-MSW oscillation hypothesis for $\tau=1.4$,
        $\tau=1.7$ and $\tau=2$.}}

\FIGURE{\epsfig{file=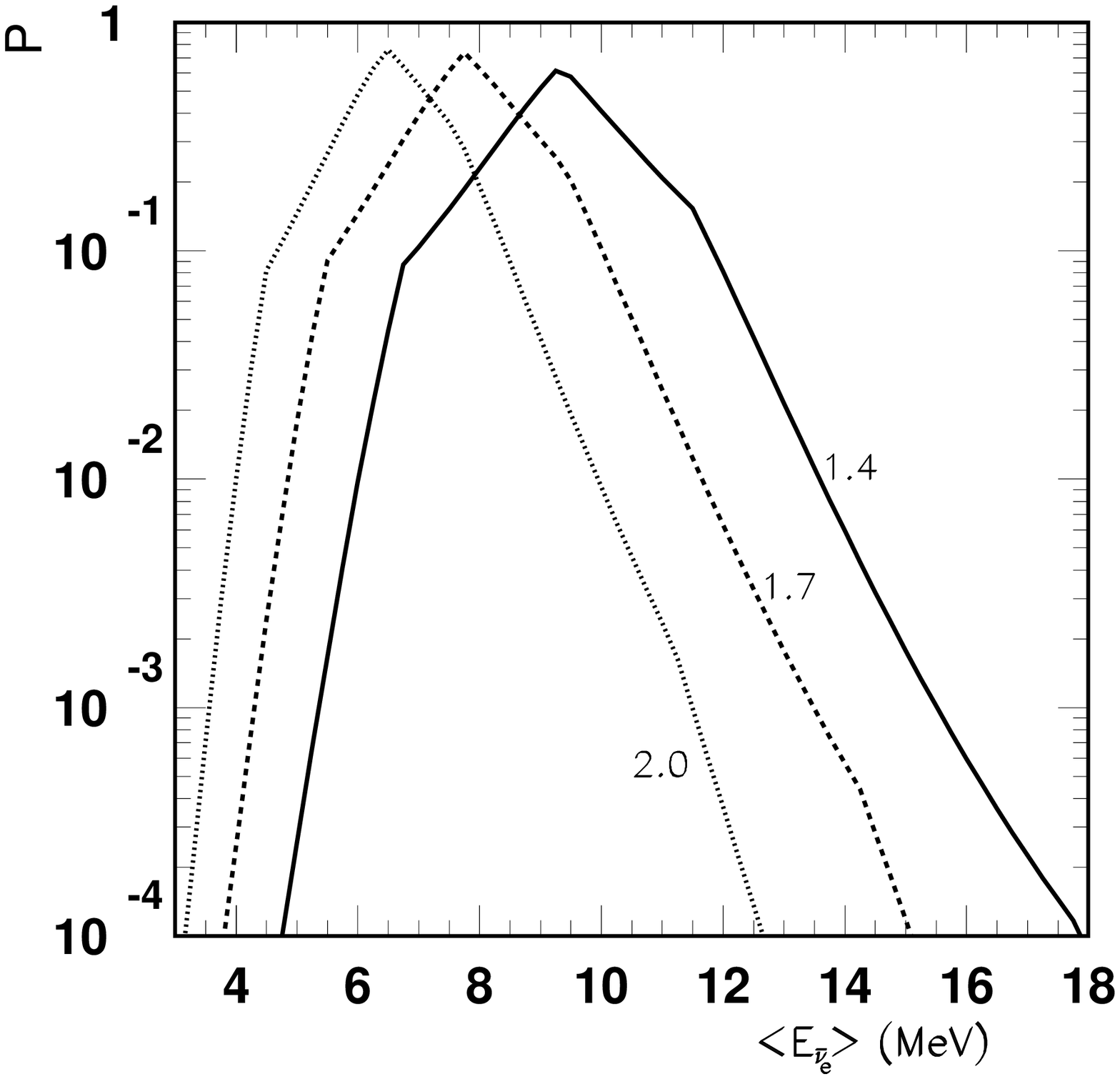,height=8.5cm,width=12.0cm}
        \caption{\label{KS1_VO}
        Probability $p$ as function of $\Ee$ given by the KS test
        that the combined data set is consistent with the VO
        hypothesis for $\tau=1.4$, $\tau=1.7$ and $\tau=2$.}} 

\section{Discussion and Conclusions}
Recently, several papers discuss the influence of neutrino
oscillations on the signal of SN~1987A.
Reference~\cite{Lu00} stresses that matter effects in the Earth
partially regenerate the $\bar\nu_e$ flux in the case of LMA-MSW
oscillations.  This effect can be clearly seen in the plots of $R$ in
the $\dm$-$\tan^2\theta$ plane\footnote{Similar figures were presented
  already in Ref.~\cite{jnr}.}, Figs.~\ref{dark_12}-\ref{dark_14},
where the likelihood of the LMA-MSW solution increases for $\dm \sim
10^{-5}$~eV$^2$.  Furthermore, Ref.~\cite{Lu00} finds that the KamII
and IMB data sets become more compatible to each other for LMA-MSW
oscillations because of the different Earth matter effects seen in the
two detectors. The general tendency of their findings agrees with ours
although there are some differences in the details. A possible
explanation for the differences is the different approach used:
Ref.~\cite{Lu00} binned the IMB data into two bins while we used a
continuous likelihood-function together with an experimental energy
resolution function.
Reference~\cite{Mi00} concentrates on the question if an inverted mass
hierarchy $m_{\nu_3}<m_{\nu_1}<m_{\nu_2}$ with $\dm_{12}=\dm_\odot$
and $\dm_{23}=\dm_{\rm atm}$ can be excluded by the SN1987 signal.
They conclude that the SN~1987A signal disfavours this scheme unless
$|U_{13}|^2\lsim$ a few $10^{-4}$.
Finally, the authors of Ref.~\cite{Mu00} consider the compatibility of
the LSND hint for $\bar\nu_\mu\leftrightarrow\bar\nu_e$ oscillations
with the SN~1987A signal and find that the LSND result is disfavoured.

In this work, we have put the main emphasis on scrutinizing whether or
not and to what extent large mixing oscillation solutions to the solar
neutrino problem are really excluded by the SN~1987A signal.  Our main
result is that the LMA-MSW solution is not in conflict with the
current understanding of supernova physics. In contrast, the VO and
the LOW solutions to the solar neutrino problem can be excluded at the
$4\sigma$ level, for most of the range of SN parameters found in
simulations.  Only a marginal region with low values of $\Ee$,
$\langle E_{\bar\nu_{\mu,\tau}} \rangle$ and $\Eb$ is left over, in
which these oscillation solutions can be reconciled with the neutrino
signal of SN~1987A.  If either of the large mixing solutions will be
finally established by future solar neutrino experiments, the neutrino
signal of SN~1987A will allow one to restrict severely the possible
range of the neutrino energies and the released binding energy in SN
explosions.

\acknowledgments
We are grateful to Andr\'e de Gouv\^ea, Hiroshi Nunokawa and Alexei
Smirnov for useful discussions. We would like to thank especially
Carlos Pe\~na for discussions and for providing the likelihood
contours of the solar neutrino data.
R.T. has been supported by a grant from the Generalitat Valenciana.
This work was also supported by Spanish DGICYT grant PB98-0693, by the
European Commission TMR networks ERBFMRXCT960090 and
HPRN-CT-2000-00148, and by the European Science Foundation network N.
86.

\end{document}